\documentclass[11pt]{amsart}
\reversemarginpar
\usepackage{amsmath}
\usepackage{amssymb}
\usepackage{graphicx}
\usepackage{graphicx}% Include figure files
\usepackage{dcolumn}% Align table columns on decimal point
\usepackage{bm}% bold math
\usepackage{amsfonts}
\usepackage{latexsym}
\usepackage{color}

\begin{document}
\newcommand{\commentout}[1]{}

\newcommand{\nwc}{\newcommand}
\newcommand{\bz}{{\mathbf z}}
\newcommand{\sqk}{\sqrt{\ks}}
\newcommand{\sqkone}{\sqrt{|\k\alpha|}}
\newcommand{\sqktwo}{\sqrt{|\k\beta|}}
\newcommand{\invsqkone}{|\k\alpha|^{-1/2}}
\newcommand{\invsqktwo}{|\k\beta|^{-1/2}}
\newcommand{\partz}{\frac{\partial}{\partial z}}
\newcommand{\grady}{\nabla_{\by}}
\newcommand{\gradp}{\nabla_{\bp}}
\newcommand{\gradx}{\nabla_{\bx}}
\newcommand{\invf}{\cF^{-1}_2}
\newcommand{\myphi}{\tilde\Theta_{(\eta,\rho)}}
\newcommand{\minrg}{|\min{(\rho,\phi^{-1})}|}
\newcommand{\al}{\alpha}
\newcommand{\xvec}{\vec{\mathbf x}}
\newcommand{\kvec}{{\vec{\mathbf k}}}
\newcommand{\lt}{\left}
\newcommand{\ksq}{\sqrt{\ks}}
\newcommand{\rt}{\right}
\newcommand{\ga}{\phi}
\newcommand{\vas}{\varepsilon}
\newcommand{\lan}{\left\langle}
\newcommand{\ran}{\right\rangle}
\newcommand{\tvas}{{W_z^\vas}}
\newcommand{\psiep}{{W_z^\vas}}
\newcommand{\wep}{{W^\vas}}
\newcommand{\weptil}{{\tilde{W}^\vas}}
\newcommand{\wepz}{{W_z^\vas}}
\newcommand{\weps}{{W_s^\ep}}
\newcommand{\wepsp}{{W_s^{\ep'}}}
\newcommand{\wepzp}{{W_z^{\vas'}}}
\newcommand{\wepztil}{{\tilde{W}_z^\vas}}
\newcommand{\vvas}{{\tilde{\ml L}_z^\vas}}
\newcommand{\veptil}{{\tilde{\ml L}_z^\vas}}
\newcommand{\vep}{{{ V}_z^\vas}}
\newcommand{\cvc}{{{\ml L}^{\ep*}_z}}
\newcommand{\cvcp}{{{\ml L}^{\ep*'}_z}}
\newcommand{\cvp}{{{\ml L}^{\ep*'}_z}}
\newcommand{\cvtil}{{\tilde{\ml L}^{\ep*}_z}}
\newcommand{\cvtilp}{{\tilde{\ml L}^{\ep*'}_z}}
\newcommand{\vtil}{{\tilde{V}^\ep_z}}
\newcommand{\ktil}{\tilde{K}}
\newcommand{\n}{\nabla}
\newcommand{\tkappa}{\tilde\kappa}
\newcommand{\Om}{{\Omega}}
\newcommand{\bx}{\mb x}
\nwc{\bv}{\mb v}
\newcommand{\br}{\mb r}
\nwc{\bH}{{\mb H}}
\newcommand{\bu}{\mathbf u}
\nwc{\bxp}{{{\mathbf x}}}
\nwc{\byp}{{{\mathbf y}}}
\newcommand{\bD}{\mathbf D}
\nwc{\bS}{\mathbf S}
\newcommand{\bA}{\mathbf \Phi}
\nwc{\cO}{\mathcal{O}}
\nwc{\co}{\mathcal{o}}
\nwc{\bG}{{\mathbf G}}
\nwc{\bF}{{\mathbf F}}
\nwc{\bd}{{\mathbf d}}
\nwc{\bR}{{\mathbf R}}
\nwc{\bh}{\mathbf h}
\nwc{\bj}{{\mb j}}
\newcommand{\bB}{\mathbf B}
\newcommand{\bC}{\mathbf C}
\newcommand{\bp}{\mathbf p}
\newcommand{\bq}{\mathbf q}
\newcommand{\by}{\mathbf y}
\nwc{\bI}{\mathbf I}
\nwc{\bP}{\mathbf P}
\nwc{\bs}{\mathbf s}
\nwc{\bX}{\mathbf X}
\newcommand{\pdg}{\bp\cdot\nabla}
\newcommand{\pdgx}{\bp\cdot\nabla_\bx}
\newcommand{\one}{1\hspace{-4.4pt}1}
\newcommand{\corr}{r_{\eta,\rho}}
\newcommand{\rinf}{r_{\eta,\infty}}
\newcommand{\rzero}{r_{0,\rho}}
\newcommand{\rzeroinf}{r_{0,\infty}}
\nwc{\om}{\omega}
\nwc{\thetatil}{{\tilde\theta}}
% theorem-like enviroments:

\nwc{\nwt}{\newtheorem}
\nwc{\xp}{{x^{\perp}}}
\nwc{\yp}{{y^{\perp}}}
\nwt{remark}{Remark}
\nwt{corollary} {Corollary}
\nwt{definition}{Definition} %def is already defined

\nwc{\ba}{{\mb a}}
\nwc{\bal}{\begin{align}}
\nwc{\be}{\begin{equation}}
\nwc{\ben}{\begin{equation*}}
\nwc{\bea}{\begin{eqnarray}}
\nwc{\beq}{\begin{eqnarray}}
\nwc{\bean}{\begin{eqnarray*}}
\nwc{\beqn}{\begin{eqnarray*}}
\nwc{\beqast}{\begin{eqnarray*}}

%\nwc{\ea}{\end{array}}
\nwc{\eal}{\end{align}}
\nwc{\ee}{\end{equation}}
\nwc{\een}{\end{equation*}}
\nwc{\eea}{\end{eqnarray}}
\nwc{\eeq}{\end{eqnarray}}
\nwc{\eean}{\end{eqnarray*}}
\nwc{\eeqn}{\end{eqnarray*}}
\nwc{\eeqast}{\end{eqnarray*}}

\nwc{\ep}{\varepsilon}
\nwc{\eps}{\varepsilon}
\nwc{\ept}{\ep }
\nwc{\vrho}{\varrho}
\nwc{\orho}{\bar\varrho}
\nwc{\ou}{\bar u}
\nwc{\vpsi}{\varpsi}
\nwc{\lamb}{\lambda}
\nwc{\Var}{{\rm Var}}

\nwt{proposition}{Proposition}
\nwt{theorem}{Theorem}
\nwt{summary}{Summary}
\nwc{\nn}{\nonumber}
%\nwc{\bm}{\boldmath}
\nwc{\mf}{\mathbf}
\nwc{\mb}{\mathbf}
\nwc{\ml}{\mathcal}

\nwc{\IA}{\mathbb{A}} %algebraic
\nwc{\bi}{\mathbf i}
\nwc{\bo}{\mathbf o}
\nwc{\IB}{\mathbb{B}}
\nwc{\IC}{\mathbb{C}} %complex
\nwc{\ID}{\mathbb{D}} %Dedekind
\nwc{\IM}{\mathbb{M}} %Dedekind
\nwc{\IP}{\mathbb{P}} %Dedekind
\nwc{\II}{\mathbb{I}} %Dedekind
\nwc{\IE}{\mathbb{E}} %Euklides
\nwc{\IF}{\mathbb{F}} %finite field
\nwc{\IG}{\mathbb{G}} %Gauss
\nwc{\IN}{\mathbb{N}} %natural
\nwc{\IQ}{\mathbb{Q}} %rational
\nwc{\IR}{\mathbb{R}} %real
\nwc{\IT}{\mathbb{T}} %torus
\nwc{\IZ}{\mathbb{Z}} %integers
\nwc{\pdfi}{{f^{\rm i}}}
\nwc{\pdfs}{{f^{\rm s}}}
\nwc{\pdfii}{{f_1^{\rm i}}}
\nwc{\pdfsi}{{f_1^{\rm s}}}
\nwc{\chis}{{\chi^{\rm s}}}
\nwc{\chii}{{\chi^{\rm i}}}
\nwc{\cE}{{\ml E}}
\nwc{\cP}{{\ml P}}
\nwc{\cQ}{{\ml Q}}
\nwc{\cL}{{\ml L}}
\nwc{\cX}{{\ml X}}
\nwc{\cW}{{\ml W}}
\nwc{\cZ}{{\ml Z}}
\nwc{\cR}{{\ml R}}
\nwc{\cV}{{\ml V}}
\nwc{\cT}{{\ml T}}
\nwc{\crV}{{\ml L}_{(\delta,\rho)}}
\nwc{\cC}{{\ml C}}
\nwc{\cA}{{\ml A}}
\nwc{\cK}{{\ml K}}
\nwc{\cB}{{\ml B}}
\nwc{\cD}{{\ml D}}
\nwc{\cF}{{\ml F}}
\nwc{\cS}{{\ml S}}
\nwc{\cM}{{\ml M}}
\nwc{\cG}{{\ml G}}
\nwc{\cH}{{\ml H}}
\nwc{\bk}{{\mb k}}
\nwc{\cbz}{\overline{\cB}_z}
\nwc{\supp}{{\hbox{supp}(\theta)}}
\nwc{\fR}{\mathfrak{R}}
\nwc{\bY}{\mathbf Y}
\newcommand{\mbr}{\mb r}
\nwc{\pft}{\cF^{-1}_2}
\nwc{\bU}{{\mb U}}

\title{%Multi-Frequency 
Compressive  Inverse Scattering II. Multi-Shot SISO Measurements with Born Scatterers}
\author{Albert C.  Fannjiang}
\thanks{The research is partially supported by
the NSF grant DMS - 0908535}. 
\email{
fannjiang@math.ucdavis.edu}

       \address{
   Department of Mathematics,
    University of California, Davis, CA 95616-8633}
   
       \begin{abstract}
Inverse scattering methods capable of compressive imaging  are 
proposed and analyzed.  The methods employ randomly and repeatedly (multiple-shot)  the single-input-single-output (SISO)
measurements  in which the probe frequencies, the incident 
and the sampling directions are  related in a precise way
and are capable of 
recovering  exactly
 scatterers of sufficiently low sparsity. 

For point targets, various sampling techniques are proposed
to transform the scattering  matrix  into
the random Fourier matrix. Two  schemes are particularly interesting:
The first one employs multiple frequencies 
with the sampling angle always in the back-scattering direction
resembling the synthetic aperture (SA) imaging; the second 
employs only single frequency with the sampling angle
in the (nearly) forward
scattering direction  in the high frequency limit,
resembling  the setting  of X-ray tomography.

The results for point targets are then extended to the case of
localized extended targets by interpolating from grid points.
In particular, an explicit error bound is derived  for 
the  piece-wise constant interpolation which 
is shown to be a practical way of discretizing localized
extended targets and enabling the compressed sensing
techniques. 

For distributed extended targets, the Littlewood-Paley basis is used 
in analysis. A specially designed  sampling scheme 
then transforms the scattering  matrix into 
a block-diagonal matrix with each block being the random
Fourier matrix corresponding to one of the multiple dyadic scales
of  the extended target. In other words by the Littlewood-Paley basis and the proposed sampling scheme the different dyadic scales of the target are decoupled and therefore  can be reconstructed scale-by-scale 
by the proposed method. Moreover,
 with probes of any single frequency $\om$ the
coefficients in the Littlewood-Paley expansion for
scales up to $\om/(2\pi)$ can be exactly  recovered.   
\commentout{An anisotropic superresolution effect is established:  In the case of anisotropic extended targets,  if  the target is comparable to wavelength in one direction,  
the target's subwavelength structures  in the other directions can be resolved.  
}

       \end{abstract}
       
       \maketitle
       
     \section{Introduction}

   %  In the high frequency limit,  diffraction tomography converges
   %  to the standard computed tomography. 
      
  \begin{figure}[t]
\begin{center}
\includegraphics[width=0.5\textwidth]{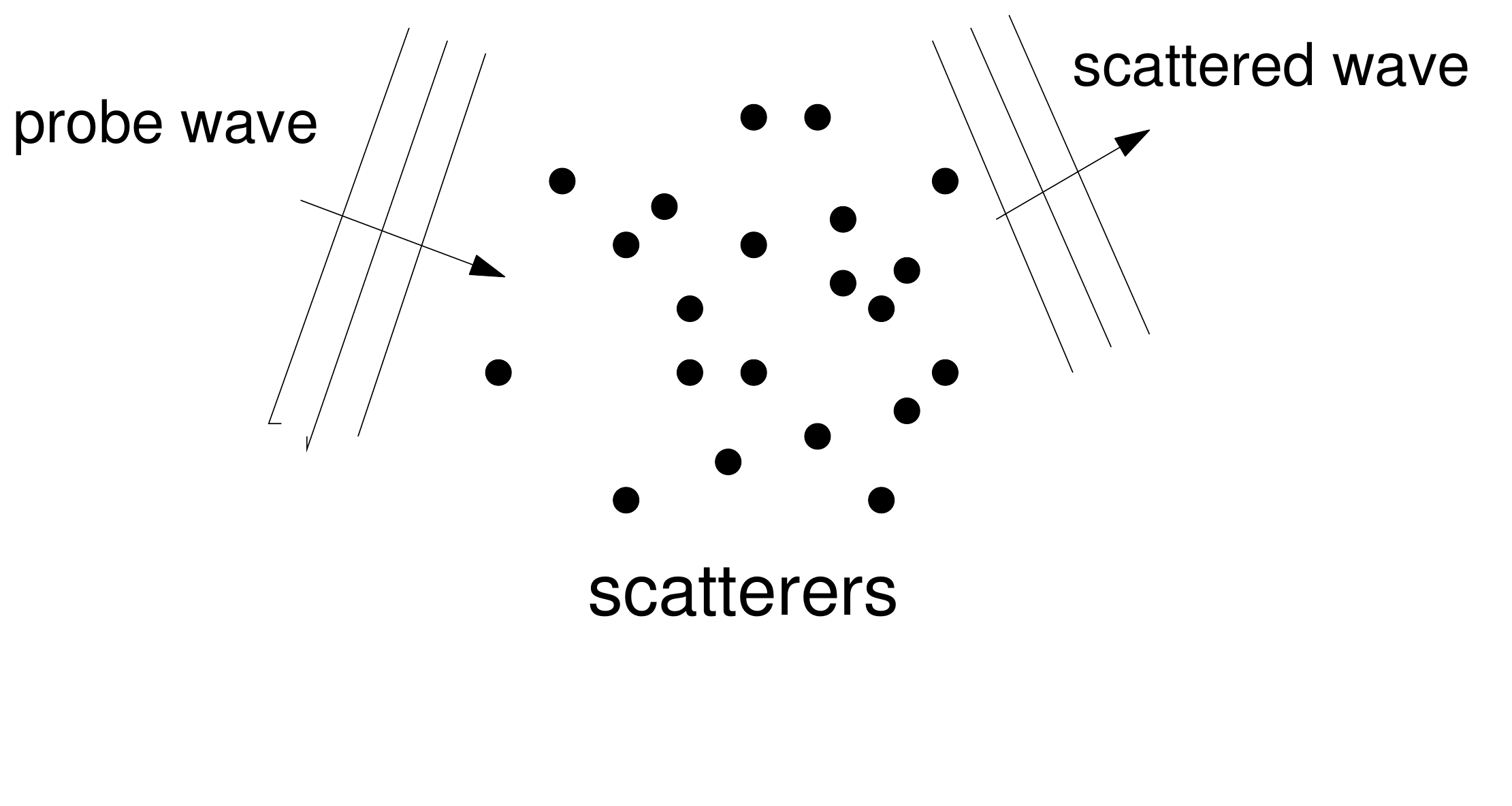}
\end{center}
\caption{Far-field  imaging of discrete or extended scatterers}
\label{fig-dt}
\end{figure}
\commentout{
\begin{figure}[t]
\begin{center}
\includegraphics[width=0.45\textwidth]{1f-figures/DT-geo.pdf}
\end{center}
\caption{Near-field imaging geometry}
\label{fig-dt}
\end{figure}
}
 
 Consider the scattering of the incident plane wave
 \beq
 u^{\rm i}(\br)=e^{i\om \br\cdot\bd}
 \eeq
 by the  variable refractive index $n^2(\br)=1+\nu(\br)$
 where $\bd$ is the incident direction. 
The scattered field  satisfies 
%\beq
%\label{scattered}
%\Delta u^{\rm s}+\om^2 u^{\rm s}=-\om^2\nu (u^{\rm i}+u^{\rm s})
%\eeq
%which can be written as
the Lippmann-Schwinger equation \cite{CK}
\beq
\label{exact'}
u^{\rm s}(\br)&=&\om^2\int \nu(\br') 
\lt(u^{\rm i}(\br')+u^{\rm s}(\br')\rt) G(\br, \br',\om)d\br',\quad\br\in \IR^d, \,\,d=2,3
\eeq
where $G(\br,\br',\om)$ is the Green function
of the operator $-(\Delta+\om^2)$. 
We assume that the wave speed is unity and hence 
the frequency equals the wavenumber $\om$.

The scattered field has
the far-field asymptotic
\beq
u^{\rm s}(\br)={e^{i\om |\br|}\over |\br|^{(d-1)/2}}\lt(A
(\hat\br,\bd,\om)+\cO(|\br|^{-1})\rt),\quad \hat\br=\br/|\br|,
\eeq
where $A$ is the scattering amplitude.
In inverse scattering theory,  
 the  scattering amplitude is the measurement data
 determined by the formula  \cite{CK}
\beq
\label{sa}
A(\hat\br,\bd,\om)&=&{\om^2\over 4\pi}
\int d\br' \nu(\br') u(\br') e^{-i\om \br'\cdot\hat\br}
\eeq

The main objective of inverse scattering then is to reconstruct
the medium  inhomogeneities  $\nu$ from the knowledge
of the scattering amplitude.  In Part I  \cite{cis-simo} and
this paper the target to be imaged consists of a finite
number of {\em point}  scatterers. And the main techniques
for reconstruction are from theory of 
{\em  compressed sensing}. In
\cite{cis-simo} we analyze the {\em one, but high, frequency} imaging method with
the single-input-multiple-output (SIMO) and multiple-input-multiple-output (MIMO) measurements
in which  for every incident plane wave the scattering amplitude
is sampled at multiple  directions {\em independent} of
the incident wave. 

In this paper the focus is on the {\em multi-shot} single-input-single-output (SISO)
measurement in which for every randomly selected  incident plane wave
the scattering amplitude is sampled  at only one
%{\em prescribed} 
 direction {\em correlated} with the incident wave. 

Our motivation for this alternative  imaging method  is practical as well as  theoretical. On the theoretical aspect,  the analysis of the high-frequency  SIMO/MIMO method 
employs  the coherence theory of compressed sensing
which 
deals with only {\em random} targets under a suitable sparsity constraint. 
On the other hand, the  {\em multi-shot} SISO
method proposed in this paper is amenable to
the restricted isometry theory of  the random Fourier matrix
which guarantees reconstruction for {\em all} targets
under the weakest known sparsity constraint. 
On the practical aspect, the present method can 
achieve a comparable  performance  with
a much lower frequency or bandwidth  (Figure \ref{fig6}). Moreover, the case of {\em extended } targets can be 
treated 
 by either interpolating from grid points or using the wavelet basis  in
 this approach. 
 The main drawback, though,  of the present approach,   in comparison
 to that of \cite{cis-simo}, 
is that  the multiple scattering effect is not accounted for.

\commentout{
  \begin{proposition} 
  Assume that $D_1$ and $D_2$ are two sound-soft scatterers
  such that their far field patterns coincide for an infinite number
  of incident plane waves with distinct directions
  and one fixed frequency. Then $D_1=D_2$.
  \end{proposition}
  If the obstacle is contained in a ball, then finitely many incident
  plane waves and the resulting far-field patterns suffice to
  determine the support of the scatterer \cite{CS, Gin}.
   A long standing problem is if the far-field pattern for one single incident direction with one fixed frequency determines
  a sound-soft scatterer without any additional {\em a priori}
  information \cite{CK}. 
  Some progress has been obtained 
  regarding uniqueness with one or two  incident plane waves
  for polyhedral scatterers  \cite{AR, CY, EY, LZ}. 
}

 In Sections  \ref{sec:mf}, we discuss the case of point scatterers
and propose several sampling schemes to transform
the scattering matrix into the random Fourier matrix
which is  amenable to the compressed sensing techniques.
One scheme  employs multiple frequencies 
with the sampling angle always in the back-scattering direction
resembling the synthetic aperture (SA) imaging; another scheme 
employs only single frequency with the sampling angle
in the (nearly) forward
scattering direction  in the high frequency limit,
resembling the setting of X-ray tomography.
We then extend these results to the case of
localized extended targets by interpolating
from grid points in Section \ref{sec:ext-loc}. 
In Section \ref{sec4}
 we analyze  the case of distributed extended targets using the Littlewood-Paley basis and propose a sampling scheme to block-diagolize 
 the scattering matrix. Each block is in the form
 of random Fourier matrix and corresponds to
 one dyadic scale of the target. Hence our method
has the capability of  imaging   the target scale-by-scale
by the compressed sensing techniques. Moreover,
 the
coefficients in the Littlewood-Paley expansion for
scales up to $\om/(2\pi)$ can be exactly  recovered
by using  probes of any single frequency $\om$.  
 We numerically test these sampling methods and compare
 their success probabilities  in  Section \ref{sec:num}. 
We  conclude and comment on the issue of resolution in Section \ref{sec:con}.

 \section{Point scatterers}
\label{sec:mf}
For the simplicity of notation, we
will focus on  two dimensions below. 

We consider the medium with point scatterers located in 
%the subset $
%\cS=\lt\{\br_{i_j}=(x_{i_j}, z_{i_j}): j=1,...,s\rt\}$
 a square lattice 
\[
\cL=\lt\{\br_i=(x_i,z_i): i=1,...,m\rt\}
\]
of spacing  $\ell$. The total number $m$ of grid points
in $\cL$  is  a perfect square. 
Without loss of generality, assume $x_j=j_1\ell, z_j=j_2\ell $
where $j=(j_1-1)\sqrt{m}+j_2$ and $j_1, j_2=1,...,\sqrt{m}$.
Let  $\nu_{j}, j=1,...,m$ be the strength of
the scatterers.  Let  $
\cS=\lt\{\br_{i_j}=(x_{i_j}, z_{i_j}): j=1,...,s\rt\}$
be the
locations of the scatterers. Hence $\nu_j=0, \forall \br_j\not \in \cS$.

For the discrete medium the scattering amplitude
becomes a finite sum 
\beq
A(\hat\br,\bd,\om)&=&{\om^2\over 4\pi}
\sum_{j=1}^m \nu_{j} u(\br_{j})
 e^{-i\om \br_j\cdot\hat\br}. 
\eeq
Unlike \cite{cis-simo} which covers both
linear and nonlinear scattering, here we
work exclusively under the Born approximation 
 in which the exciting
   field $u(\br_{j})$  is replaced 
   by the incident field $u^{\rm i}(\br_{j})$;
   unlike \cite{cis-simo} which deals  exclusively
 with  one frequency, here we will work with
 multiple frequencies.  
   
Let $\bd_l, \hat\br_l, l=1,...,n$ be various incident and sampling directions for the frequencies $\om_l, l=1,...,n$ to be determined later.
Define  the measurement vector $Y=(Y_l)\in \IC^n$ 
with
\beq
\label{4'}
Y_l={4\pi\over \om^2}A(\hat\br_l,\bd_l,\om_l),\quad l=1,...,n. 
\eeq
  The measurement vector is related to the target
vector $X=(\nu_j)\in \IC^m$  by the sensing matrix  $\bA$ as 
  $Y=\bA X$.   
  Let  $\theta_l,\tilde\theta_l$ be the polar angles of
  $\bd_l,\hat\br_l$, respectively. 
  The $(l,j)$-entry of $\bA\in \IC^{n\times m}$ is
   \beq
   \label{entry}
e^{-i\om_l\hat\br_l\cdot\br_j}e^{i\om_l\bd_l\cdot \br_j}&=&
e^{i\om_l\ell (j_2(\sin{\theta_l}-\sin{\thetatil_l})+j_1(\cos{ \theta_l }-\cos{\thetatil_l}))},\quad j=(j_1-1)+j_2. 
\eeq
Note the all entries of $\bA$ have unit modulus. 
As in \cite{cis-simo} we  reconstruct $X$ as the solution
of   the $L^1$-minimization, called
 Basis Pursuit (BP):
\beq
\min \|Z\|_1 \qquad \text{s.t.} \,\, \bA Z=Y  
\label{L1}
\eeq
which can be solved by linear  program or by  various  greedy
algorithms \cite{BDE, CDS, Tem}. 
\commentout{
In the presence of noise/error
 of size $\ep$, (\ref{L1}) is replaced by 
 \beq
 \min \|X\|_1 \qquad \text{s.t.} \,\, \|\bA X-Y\|_2\leq \ep.   \label{L1'}
\eeq
}

In the presence of noise/error $E$ of size $\ep$ as
in  
\beq
\label{31}
Y=\bA X+E,\quad \|E\|_2\leq \ep
\eeq
 (\ref{L1}) is replaced by 
the relaxation scheme
\beq
\label{32}
\mbox{min}\,\, \|Z\|_1,\quad \mbox{s.t.}\quad \|Y-\bA Z\|_2\leq \epsilon.
\eeq
%with some constant $c>0$.

A fundamental notion in compressed sensing under which
BP yields the unique exact solution is the restrictive isometry property (RIP) due to Cand\`es and Tao \cite{CT}.
Precisely, let  the sparsity $s$  of a vector  $Z\in \IC^m$ be the
number of nonzero components of $Z$ and define the restricted isometry constant $\delta_s$
to be the smallest positive number such that the inequality
\[
\kappa (1-\delta_s) \|Z\|_2^2\leq \|\bA Z\|_2^2\leq \kappa (1+\delta_s)
\|Z\|_2^2
\]
holds for all $Z\in \IC^m$ of sparsity at most $ s$ and some
constant $\kappa>0$. 
For the target vector $X$ let $X^{(s)}$ denote the best $s$-sparse
approximation of $X$ in the sense of $L^1$-norm, i.e.
\[
X^{(s)}=\hbox{argmin}\,\, \|Z-X\|_1,\quad\hbox{s.t.}
\quad \|Z\|_0\leq s
\]
where $\|Z\|_0$ denotes the number of nonzero components, called the sparsity,  of $Z$. 
Clearly, $X^{(s)}$ consists of the $s$ largest components
of $X$.

Now we state the fundamental   result 
of the RIP approach \cite{Can} which is an improvement
of the results of \cite{CRT1, CT}. 
\begin{proposition} \label{rip} \cite{Can}
Suppose the  restricted isometry constant of $\bA$
satisfies the inequality 
\beq
\label{ric}
\delta_{2s}<\sqrt{2}-1
\eeq
with $\kappa=1$. 
 Then the solution $\hat X$ by BP (\ref{32}) satisfies
 \beq
% \label{100}
% \|\hat X-X\|_1&\leq & C_0\|X-X^{(s)}\|_1\\
 \|\hat X-X\|_2&\leq & C_1s^{-1/2}\|X-X^{(s)}\|_1+C_2\ep
 \eeq
 for some constants $C_1$ and $C_2$. 
 %In particular,
 %if $X$ is $s$-sparse, then the recovery is exact.
\end{proposition}
\begin{remark}
\label{rmk1}
For general $\kappa\neq 1$, one can consider the normalized
version of (\ref{31}) 
\[
{1\over \sqrt{\kappa}}Y={1\over \sqrt{\kappa}}\bA X+
{1\over \sqrt{\kappa}}E
\]
by which it follows that
\beq
 \|\hat X-X\|_2&\leq & C_1s^{-1/2}\|X-X^{(s)}\|_1+C_2{\ep\over\sqrt{\kappa}}\label{101}. 
 \eeq

\end{remark}
\commentout{
\begin{remark}
\label{rmk:sp}
Greedy algorithms have significantly lower  computational
complexity than linear programming and have 
provable performance under various conditions.
For example
under the condition  $\delta_{3s}<0.06$
 the Subspace Pursuit (SP) algorithm is guaranteed to exactly recover $X$ via a finite number of iterations \cite{DM}.
 
\end{remark}
}

We wish to write the $(l,j)$-entry of the sensing matrix in the form
\beq
\label{111}
e^{i \pi (j_1\xi_l+j_2\zeta_l)},\quad j=(j_1-1)\sqrt{m}+j_2,\quad j_1, j_2=1,...,\sqrt{m},\quad l=1,...,n
\eeq
where $\xi_l, \zeta_l$ are independently and
uniformly distributed in $[-1,1]$ in view of
the following theorem.

\begin{proposition}\cite{Rau}\label{prop2}
Suppose \beq
\label{77}
{n\over \ln{n}}\geq C \delta^{-2}\sigma\ln^2{\sigma} \ln{m} \ln{1\over \alpha},\quad
\alpha\in (0,1)
\eeq
for a given sparsity $\sigma$ where $C$ is an absolute constant. 
Then the restricted isometry constant 
of  the matrix with entry (\ref{111}) satisfies 
\[
\delta_\sigma\leq \delta
\]
 with $\kappa=n$ and with
probability at least $1-\alpha$.
\end{proposition}
See \cite{CRT1, CT2,  RV}  for the case when $\xi_l,\zeta_l$ belong to 
the discrete  subset of  $[-1,1] $ of equal spacing $2/\sqrt{m}$.

To construct a sensing matrix of the form (\ref{111})
we proceed as follows.  
Write $(\xi_l,\zeta_l)$ in the polar coordinates $\rho_l, \phi_l$ as 
\beq
(\xi_l,\zeta_l)=\rho_l (\cos\phi_l,\sin\phi_l),\quad 
\rho_l=\sqrt{\xi_l^2+\zeta_l^2} \leq\sqrt{2}
\eeq
and set 
\beqn
\om_l(\cos\theta_l-\cos\tilde\theta_l)& =&\sqrt{2}\rho_l\Om\cos\phi_l\\
\om_l(\sin\theta_l-\sin\tilde\theta_l)&=&\sqrt{2}{\rho_l\Om } 
\sin \phi_l
\eeqn
where $\Om$ is a parameter to be determined later (\ref{75}). 
Equivalently we have
\beq
\label{200}
-\sqrt{2}\om_l\sin{\theta_l-\tilde\theta_l\over 2}\sin{\theta_l+\tilde\theta_l\over 2}& =& \Omega\rho_l\cos\phi_l\\
\sqrt{2} \om_l\sin{\theta_l-\tilde\theta_l\over 2}\cos{\theta_l+\tilde\theta_l\over 2}&=&\Omega\rho_l\sin \phi_l. 
\label{201}
\eeq
This set of equations determines the single-input-$(\theta_l,\om_l)$-single-output-$\tilde\theta_l$ 
mode of sampling.

The following   implementation of  (\ref{200})-(\ref{201}) is natural.  Let the sampling angle $\tilde\theta_l$ be related
to the incident  angle $\theta_l$ via 
\beq
\label{ch1}
\theta_l+\tilde\theta_l=2\phi_l+\pi,%\quad\theta_l,\tilde\theta_l\in [0,2\pi]
\eeq
%such that $\theta_l-\tilde\theta_l\in [0,2\pi]$ 
and set the frequency $\om_l$ to be
\beq
\label{ch2}
\om_l= {\Omega \rho_l\over \sqrt{2} \sin{\theta_l-\tilde\theta_l\over 2}}.
\eeq
Then the entries (\ref{entry}) of the sensing matrix $\bA$ 
have the form
\beq
e^{i\sqrt{2}\Om \ell(j_1\xi_l+j_2\zeta_l)}, \quad
l=1,...,n,\quad j_1, j_2=1,...,\sqrt{m}.\label{17-2}
\eeq
By the square-symmetry of the problem,
it is clear that the relation (\ref{ch1}) can be generalized to 
\beq
\label{ch1'}
\theta_l+\tilde\theta_l=2\phi_l+\eta \pi,\quad \eta\in \IZ.   
\eeq
On the other hand, the symmetry of the square lattice 
should not  play a significant role and
hence we expect the result to be 
insensitive to 
any {\em fixed} $\eta\in \IR$, independent of $l$, as
long as (\ref{ch2}) holds.
Indeed this is confirmed by numerical simulations below 
(Figure  \ref{fig4}).

\commentout{
Alternatively 
\[
\theta_l+\tilde\theta_l=2\phi_l-\pi,\quad\theta_l,\tilde\theta_l\in [0,2\pi]
\]
such that $\tilde\theta_l-\theta_l\in [0,2\pi]$ and
\[
\om_l= {\Omega\over \sin{\tilde\theta_l-\theta_l\over 2}}\sqrt{\xi_l^2+\zeta_l^2\over 2}\in [0,\Om].
\]
}

Let us focus on  three specific measurement schemes.

\bigskip

\noindent{\bf Scheme  I.} 
This scheme employs  $\Om-$band limited probes, i.e.
$\om_l\in  [-\Om,\Om]$. 
This and (\ref{ch2}) lead to
the constraint:
\beq
\label{const}
{\lt| \sin{\theta_l-\tilde\theta_l\over 2}\rt|}\geq {\rho_l\over\sqrt{2}}. 
\eeq

The simplest way  to satisfy (\ref{ch1}) and (\ref{const}) is
to set 
\beq
\label{41}\phi_l&=&\tilde\theta_l=\theta_l+\pi,\\
\label{42} \om_l&=& {\Om\rho_l\over \sqrt{2}}
\eeq
$l=1,...,n$. In this case the scattering amplitude is always sampled in 
the back-scattering direction. This resembles
the synthetic aperture imaging which has been 
previously analyzed under the paraxial approximation
in \cite{cs-par}. 
 In contrast, the forward
scattering direction with $\tilde\theta_l=\theta_l$ almost surely violates
the constraint  (\ref{const}). 

\bigskip

\noindent {\bf Scheme II.}
 This scheme employs  single  frequency probes no less
 than $\Omega$: 
\beq
\label{20}
\om_l=\gamma\Omega,\quad \gamma\geq 1,\quad   l=1,...,n.\eeq
To satisfy (\ref{ch1'}) and (\ref{ch2}) we set 
\commentout{
and set
\[
\sin{\theta_l-\tilde\theta_l\over 2}={\rho_l\over \gamma \sqrt{2}}
\]
which is solved by
}
\beq
\label{21}
\theta_l=\phi_l+{\eta\pi\over 2} +\arcsin{\rho_l\over \gamma\sqrt{2}}\\
\tilde\theta_l=\phi_l+{\eta\pi\over 2}-\arcsin{\rho_l\over \gamma\sqrt{2}}\label{22}
\eeq
with $\eta\in \IZ$. 
The difference between the incident angle and
the sampling angle is 
\beq
\label{25}
\theta_l-\tilde\theta_l=2\arcsin{\rho_l\over \gamma\sqrt{2}}
\eeq
which diminishes as $\gamma\to\infty$. In other words, in the high frequency limit, the sampling angle approaches the
incident  angle. This resembles the setting of  the X-ray tomography.

\commentout{
Let 
\[
g(\om)={2\om\over \Omega^2},\quad \om\in (0, \Omega)
\]
and 
$\pdfs(\thetatil) $ the uniform distribution in $[-\pi,\pi]$.
As a result, $(\om \cos{\thetatil}, \om\sin{\thetatil})$
is uniformly distributed in the circular disk of radius
$\Omega$ containing the square $(-\Omega/\sqrt{2}, 
\Omega/\sqrt{2})^2$. 
} 
\commentout{
Hence $(\om \cos{\thetatil}, \om\sin{\thetatil})$
is in the circular disk of radius
$\Omega$ containing the square $(-\Omega/\sqrt{2}, 
\Omega/\sqrt{2})^2$. Now consider
the finite sequence of independently and uniformly distributed
random vectors $\bv_j \in (-\Omega/\sqrt{2}, 
\Omega/\sqrt{2})^2, j=1,...,n.$
%We shall restrict the measurement to 
%$(\om \cos{\thetatil}, \om\sin{\thetatil})\in(-\Omega/\sqrt{2}, \Omega/\sqrt{2})^2 $
%by proper conditioning. 
Let $\bv_l=\omega_l(\cos\thetatil_l,\sin\thetatil_l)$ with $ \omega_l=|\bv_l|\in (0,\Omega)$.  
By design, the sensing
matrix $\bA$ has the entries
\[
e^{i\Omega \ell (j_1\xi_l+j_2\zeta_l)},\quad j_1, j_2=1,...,\sqrt{m},\quad l=1,...,n
\]
where $\xi_l, \zeta_l$ are independently and
uniformly distributed in $[-1,1]$. 
}

\bigskip

\noindent {\bf Scheme III.}
This scheme employs probes of unlimited frequency band.
Let $\theta_l$ be $n$ arbitrary distinct numbers in $ [-\pi,\pi]$ 
and let $\tilde\theta_l$ and $\om_l$ be determined
by (\ref{ch1'}) and (\ref{ch2}), respectively. 
The possibility of having a small divisor in (\ref{ch2})
renders  the bandwidth  unlimited in principle.

The 
following  result  is an immediate consequence 
of   Proposition \ref{rip} 
and Proposition \ref{prop2}. 

\begin{theorem} \label{thm:mf}
Let $\xi_l, \zeta_l$ be independently and
uniformly distributed in $[-1,1]$ and 
let $(\rho_l,\phi_l)$ be the polar coordinates
of  $(\xi_l,\zeta_l)$, i.e. 
\[
(\xi_l,\zeta_l)=\rho_l(\cos\phi_l,\sin\phi_l).
\]

Let the probe frequencies $\om_l$, the incident angles $\theta_l$
and the sampling angles $\tilde\theta_l$ satisfy (\ref{ch1})
and (\ref{ch2}),  for example, by Scheme I, II or III. 
%imposing  (\ref{41})-(\ref{42}) or (\ref{20})-(\ref{22}).
 
Suppose 
\beq
\label{75}
\Omega\ell=\pi/\sqrt{2}
\eeq
and suppose (\ref{77}) holds with $\sigma=2s$ and any $\delta<\sqrt{2}-1$.   Then (\ref{ric})  with $\kappa=n$ is satisfied 
for the matrix $\bA$ 
and the bound (\ref{101})
holds  true with probability at least $1-\alpha$.
%In particular, every target vector of sparsity less than $s$
%can be exactly recovered  by  BP (\ref{L1}).
\end{theorem}

\commentout{
Next, let us consider the case with an arbitrary (fixed) set of
 incidence angles
$\theta_k, k=1,...,n$.
\commentout{ 
Note that the random vector 
\[
\bv\equiv (\cos{\theta}-\cos{\thetatil}, \sin{ \theta}-\sin{\thetatil})
\]
has a  circularly symmetric probability density function 
and its magnitude 
\[
|\bv|
=2(1-\sin{(\theta+\thetatil)})\in [0,4]
\]
has the probability density function 
\[
{1\over \sqrt{16-v^2}},\quad v=|\bv|
\]
Hence we may write
\[
\bv=v(\cos\phi,\sin\phi)
\]
where $\phi$ is uniformly distributed in $[0,2\pi]$. 
}
Note that
\[
\lt|(\cos{\theta}-\cos{\thetatil}, \sin{ \theta}-\sin{\thetatil})\rt|
=\sqrt{2(1-\sin{(\theta+\thetatil)})}\in [0,2].
\]
Let  $\bv_l, l=1,...,n,$ be a sequence of independently and uniformly
distributed  random vectors in $(-\sqrt{2}\Omega, \sqrt{2}\Omega)^2$.  Set
\[
\bv_l=2\omega_l (\cos\phi_l, \sin\phi_l)
\]
with $|\bv_l|=2\omega_l$ and
let
\[
\thetatil_l=\pi+\theta_l-2\phi_l,\quad l=1,...,n
\]
be the angles for collecting the data.

Analogous to Theorem \ref{thm:mf}, we have the following theorem.

\begin{theorem} \label{thm:mf2}
If 
\beq
\label{76-2}
\Omega\ell={\pi\over {2}}
\eeq
and 
\[
{w\over \ln{w}}\geq C s\ln^2{s} \ln{m} \ln{1\over \alpha}
\]
for $\alpha\in (0,1)$ and some absolute constant $C$,  then the restricted isometry condition (\ref{ric}) is satisfied 
with probability at least $1-\alpha$
and hence all target amplitudes of sparsity less than $s$
can be uniquely determined from BP (\ref{L1}).
\end{theorem}

In both (\ref{75}) and (\ref{76-2}), the resolution $\ell$ is inversely proportional to
the bandwidth $\Omega$. 
Notice, however,  the two-fold enhancement of the resolving power expressed by
(\ref{76-2}) over that by (\ref{75}), which has been
previously established in a different context
\cite{subwave-cs}. The superiority of Theorem \ref{thm:mf}
and \ref{thm:mf2} over Theorem \ref{thm1} is that
the exact recovery holds for {\bf all} targets
under  condition (\ref{75}) and (\ref{76-2}), respectively.
Moreover, the sparsity constraint (\ref{77}) is weaker
than (\ref{spark}). 
}

\commentout{
The next result  is a restatement of the result of \cite{Can} 
after applying  Proposition  \ref{prop2} with $\sigma=2s, \delta<\sqrt{2}-1$. 
\begin{theorem}\label{thm8}
Let $Y$ be given by (\ref{31}) and  let $\hat X$ be the solution to (\ref{32}).
Then under the assumptions of Theorem 1 
we have 
 \beq
 \|\hat X-X\|_2\leq C_1s^{-1/2} \|X-X^{(s)}\|_1+ C_2\ep
 \eeq
with
probability at least $1-\alpha$
where  $C_1$ and $C_2$ are
constants.   \end{theorem}
}

\commentout{
\begin{proposition}\cite{CRT2}
 Let $\hat X$ be the solution to (\ref{32}).
 If  $\|X\|_0\leq s$ and (\ref{ric}) is true, then
 \[
 \|\hat X-X\|_2\leq C_s \ep
 \]
 where $C_s$ depends only on  and behaves reasonably with $\delta_{4s}$. 
 \end{proposition}
 }

 \section{Localized  extended targets}\label{sec:ext-loc}
 
  \begin{figure}[t]
\begin{center}
\includegraphics[width=0.45\textwidth]{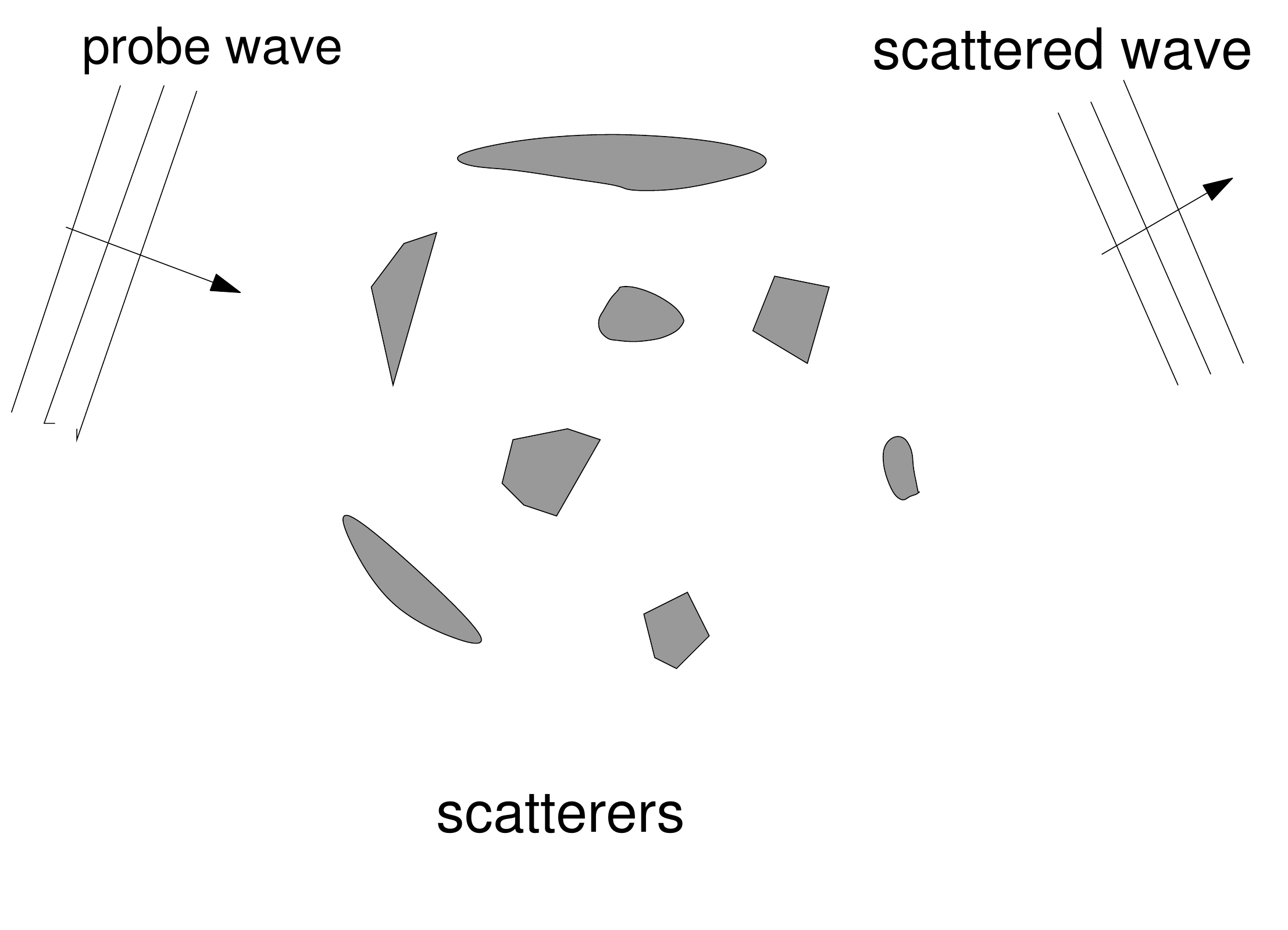}
\end{center}
\caption{Scattering from extended targets}
\end{figure}

In this section we consider  extended targets that
are localized in space. We 
represent such targets by interpolating from grid points
and thus extend the preceding results for point targets
to localized extended targets. We
will treat the case of distributed extended targets
that are not localized in space in the next section.

Suppose that the target function $\nu(\br)$ is continuous 
and has  a compact support. 
Consider the filtered version $\nu_\eta$ of  $\nu$
\beq
\label{531}
\nu_\eta(\br)=\int g_\eta(\br-\br') \nu(\br') d\br'
\eeq
where  $g_\eta(\br)=
\eta^{-2} g(\br/\eta)$ for any integrable filter function $g$ such that $\int g(\br) d\br =1$. 
Clearly, $\nu_\eta$ tends to $\nu$ as  $\eta\to 0$.
%Clearly, $\|v_\eta-v\|_\infty\to 0$ as $\eta\to 0$. Moreover,
%for square-integrable $v$, $v_\eta$ tends to $v$ in $L^2$
%as $\eta \to 0$.  The size of $\eta$ will be determined later.

Next we discretize (\ref{531}) by replacing the integral by
the Riemann sum of step size $\ell$. We obtain
\beq
\label{533}
\nu_{\eta,\ell}(\br)=\ell^2\sum_{\bp\in I}
g_\eta(\br-\ell\bq ) \nu(\ell\bq),\quad I\subset \IZ^2 
\eeq
which is as smooth as the interpolation element $g_\eta$ is. 
%For sufficiently small $\eta, \ell$,
%$\|v_{\eta, \ell}-v\|_{1}$ can be made arbitrarily small. 
\commentout{ the difference in
the Sobolev norm $\|v_{\eta, \ell}-v\|_{k,1}$ can be
made arbitrarily small where
\[
 \|v\|_{k,1}=\max_{j\leq k} \lt\|{\partial^k\over \partial x^j\partial z^{k-j}} v \rt\|_1,\quad k\in \IN. 
 \]
 }

Since $\nu$ has a  compact support, $I$ is a finite set. 
For simplicity let  $I$ be the square sublattice 
\[
I=\{\bq=(q_1, q_2): q_1, q_2= 1,...,\sqrt{m}\}
\]
of total cardinality $m$. Let $j=(q_1-1)\sqrt{m}+ q_2$. 
Define the target vector $X=(X_j)\in \IC^m$ with $X_j=\nu(\ell\bq)$. Let $\om_l$ and $ \bd_l$ be the probe frequencies
and directions, respectively, and let $\hat\br_l$ be the
sampling directions for $l=1,...,n$. 
Now we write
the data vector $Y$ 
in the form (\ref{31}) with  the sensing matrix elements 
\beq
\label{535}
\Phi_{lj}&=&{1\over  2\pi  \hat g(\eta\om_l(\bd_l-\hat\br_l))}
\int_{\IR^2} g_\eta(\br'-\ell\bq) e^{i\om_l (\bd_l-\hat\br_l)\cdot \br'} d\br',\quad j=(q_1-1)\sqrt{m}+ q_2\\
&=& e^{i\om_l \ell \bq\cdot (\bd_l-\hat\br_l)}\nn
\eeq
and the error term $E$ due to the filtered discretization. 
For sufficiently small $\eta, \ell$ we may assume $\|E\|_2\leq \ep n^{1/2}$ for a given $\ep>0$ (see Remark \ref{rmk1} and below). 

The crucial  observation is that the sensing matrix (\ref{535}) is
identical to (\ref{entry}) with $(x_j, z_j)=\om\bq$ and
any {\em isotropic} filter function $g$.
Therefore
Theorem \ref{thm:mf} holds  verbatim  for
 the case of localized  extended targets formulated
above. 

How small must $\eta$ and $\ell$ be in order to ensure
that $\|E\|_2\leq \ep n^{1/2}$? This  
can be answered roughly as follows. First, by the inequality $\|E\|_2\leq \|E\|_\infty \sqrt{n}$
it suffices to have $\|E\|_\infty \leq \ep $. Now
consider the transformation
$\cT$, defined by 
\[
 (\cT \nu)_l ={1\over 2\pi \hat g(\eta\om_l(\bd_l-\hat\br_l))} \int \nu(\br') e^{i\om_l (\bd_l-\hat \br_l ) \cdot \br'} d\br', 
 \]
 from  the space of continuous functions supported on  $[\ell, m\ell]^2$ to
$\IC^n$,  cf. (\ref{sa}). By definition
  \[
 E= \cT v -\cT v_{\eta,\ell}
 \]
 and we have
 \[
 \|E\|_\infty\leq {\|\nu-\nu_{\eta,\ell}\|_1\over  2\pi \min_{l}|\hat g(\eta\om_l(\bd_l-\hat\br_l))|}.
 \]
 
 %\[
%g_\eta(\br)={1\over {2\pi}\eta^2} e^{-{|\br|^2\over 2\eta^2}}. 
%\]
For  the sampling schemes I and II (with $\gamma=1$)  of Section \ref{sec:mf}, we can give an explicit bound for
the discretization error
\beqn
\|E\|_\infty\leq {1\over 2\pi}  {\|\nu-\nu_{\eta,\ell}\|_1}
\|\hat g^{-1}\|_{L^\infty([-2\eta\Om, 2\eta \Om]^2)}. 
\eeqn
where $\|\cdot\|_{L^\infty([-2\eta\Om,  2\eta\Om]^2)}$ denotes
the $L^\infty$-norm of functions defined on $[-2\eta\Om,  2\eta\Om]^2$. 
 
 \begin{theorem} \label{thm:loc-ext}
Consider the sampling schemes I and II (with $\gamma=1$)  of Section \ref{sec:mf}. Assume  (\ref{75})  and
\beq
\label{disc-err}
 {\|\nu-\nu_{\eta,\ell}\|_1}
\leq  {2\pi \ep \over \|\hat g^{-1}\|_{L^\infty([-2\eta\Om,  2\eta\Om]^2)}}. 
\eeq
Suppose that (\ref{77}) 
 holds with $\sigma=2s$ and any $\delta<\sqrt{2}-1$.   Then 
the bound (\ref{101}) with $\kappa=n$
holds  true  with probability at least $1-\alpha$.
%In particular, every target vector of sparsity less than $s$
%can be exactly recovered  by  BP (\ref{L1}).
\end{theorem}

\begin{remark}
For example, consider the smooth isotropic filter function 
 \[
g(\br)={1\over {2\pi}} e^{-{|\br|^2\over 2}} 
\]
with
$
\hat g(\bk)=g(\bk).$ 
With the choice $\eta=\ell$, the condition (\ref{disc-err}) can be formulated as
\beq
\label{gau}
  {\|\nu-\nu_{\ell,\ell}\|_1}\leq
 e^{\pi^2} \ep. 
 \eeq
 
 For another example, consider the indicator function  $g$ on
 the unit square $[-{1\over 2}, {1\over 2}]^2$. The resulting
 interpolation  is the
 piece-wise constant approximation. Then
 \[
 \hat g(k_1, k_2)={2\over \pi}\cdot {\sin{k_1\over 2}\over k_1} \cdot
 {\sin{k_2\over 2}\over k_2}
 \]
 Setting  again $\eta=\ell$ we obtain the condition
 \beq
 \label{piece}
  {\|\nu-\nu_{\ell,\ell}\|_1}\leq
 {\pi^2 \ep\over 2\sin^2(\pi/\sqrt{2})}
 \eeq
 which  appears to be  much more useful than (\ref{gau}).

\end{remark}

 \section{Distributed extended targets}\label{sec4}

In this section, we consider  extended targets
represented by  square-integrable functions $\nu(x,z)$.

To this end
 we 
  use  the Littlewood-Paley basis
 \beq
 \label{lp}
 \hat \psi(\xi,\zeta)=\lt\{
 \begin{matrix}
 (2\pi)^{-1}. & \pi\leq |\xi|, |\zeta|\leq 2\pi\\
 0,&\hbox{otherwise}
 \end{matrix}
 \rt.
 \eeq
 or
 \beq
 \label{lp2}
 \psi(\br)=(\pi^2 xz)^{-1} (\sin{(2\pi x)}-\sin{(\pi x)})\cdot
(\sin{(2\pi z)}-\sin{(\pi z)}). 
 \eeq
 Then the following set of functions 
 \beq
 \psi_{\bp,\bq}(\br)=2^{-(p_1+p_2)/2}\psi(2^{-\bp}\br-\bq),\quad \bp,\bq\in \IZ^2
 \eeq
 with 
 \[
 2^{-\bp}\br=(2^{-p_1}x,2^{-p_2}z)
 \]
 forms an orthonormal wavelet basis in $L^2(\IR^2)$ \cite{Dau}. 
In terms of the Littlewood-Paley basis we have  the expansion 
 \beq
 \label{30}
 \nu(x, z)=\sum_{\bp,\bq\in \IZ^2}\nu_{\bp,\bq} \psi_{\bp,\bq}(x,z).
 \eeq

With the incident fields
\[
u_k^{\rm i}(\br)=e^{i\om_k \br\cdot\bd_k},\quad k=1,...,n
\]
we have from  (\ref{sa}), (\ref{4'})
and (\ref{30}) that 
\beq
\label{7}
Y_k={ 2\pi} \sum_{\bp,\bq\in \IZ^2} 2^{(p_1+p_2)/2} \nu_{\bp,\bq} e^{i\om_k 2^\bp(\bd_k-\hat\br_k)\cdot \bq}\hat\psi(\om_k 2^\bp(\hat\br_k-\bd_k)),\quad k=1,...,n. 
\eeq

 Define
 the sensing matrix elements to be
\beq
\label{sense22}
\Phi_{k,l}= {1\over {2n_\bp+1}}\hat\psi(\om_k 2^\bp
(\hat\br_k-\bd_k))
 e^{i\om_k 2^{\bp}(\bd_k-\hat\br_k)\cdot \bq}
\eeq
and let  $\bA=[\Phi_{k, l}]$, where
$\bd_k,\hat\br_k,\om_k$ are given below.

 Let 
 \beqn
l&=&\sum_{j_1=-p_*}^{p_1-1}\sum_{j_2=-p_*}^{p_2-1}(2m_{\bj}+1)^2+(q_1+m_{\bp})(2m_{\bp}+1)+(q_2+m_\bp+1),\quad |\bq|_\infty\leq m_\bp, \quad |\bp|_\infty\leq p_*, \\
k&=&\sum_{j_1=-p_*}^{p'_1-1}\sum_{j_2=-p_*}^{p'_2-1}(2n_{\bj}+1)^2+(q'_1+n_{\bp'})(2n_{\bp'}+1)+(q'_2+n_{\bp'}+1),\quad |\bq'|_\infty\leq n_{\bp'},
\quad |\bp'|_\infty\leq p_*, 
\eeqn
 for some $m_\bp, n_\bp, p_*\in \IN$, be the column and row indexes, respectively, of $\bA$. It is important to keep in mind
 how  $k$ and $l$ are related to
 $(\bp',\bq')$ and $(\bp,\bq)$, respectively, in order to 
 understand the scheme described below. 
%Let $n=\sum_{|\bp|_\infty\leq p_*}(2n_{\bp}+1)^2$. 
 
 Let $\xi_k, \zeta_k$ be independent, uniform random variables on 
$[-1,1]$ and define
\beq
\label{40}
\alpha_k={\pi \over \om_k 2^{p_1'}}
\cdot\lt\{\begin{matrix}
1+\xi_k,& \xi_k\in [0,1]\\
-1+\xi_k,&\xi_k\in [-1,0]
\end{matrix}\rt.\\
\label{40'}
\beta_k={\pi \over \om_k 2^{p_2'}}
\cdot\lt\{\begin{matrix}
1+\zeta_k,& \zeta_k\in [0,1]\\
-1+\zeta_k,&\zeta_k\in [-1,0]
\end{matrix}\rt..
\eeq
Suppose $\alpha_k,\beta_k \in  [-1,1]$ for all $\bp', |\bp'|
\leq p_*$. This holds true, for example,  when  the frequencies 
 $\om_k$ satisfy the constraint 
\beq
\label{freq}
\om_k 2^{p_1'}\geq 2\pi,\quad \om_k 2^{p_2'}\geq {2}\pi. 
 \eeq
Let $(\rho_k,\phi_k)$
be the polar coordinates of $(\alpha_k,\beta_k)$.

As before, let $\theta_k, \tilde\theta_k$ be the angles
of incidence and sampling, respectively,  which
are chosen according to 
%Let (\ref{200})-(\ref{201}) with $\Omega=\pi/\sqrt{2}$  be satisfied, namely
\beq
\label{200'}
-{2}\sin{\theta_k-\tilde\theta_k\over 2}\sin{\theta_k+\tilde\theta_k\over 2}& =& \alpha_k= \rho_k\cos\phi_k\\
{2} \sin{\theta_k-\tilde\theta_k\over 2}\cos{\theta_k+\tilde\theta_k\over 2}&=&\beta_k=\rho_k\sin \phi_k
\label{201'}
\eeq
in analogy to  (\ref{200})-(\ref{201}). 
This means
\beq
\label{support}
-\om_k 2^{\bp}(\hat\br_k-\bd_k)=\om_k(2^{p_1}\alpha_k, 2^{p_2}\beta_k)=\pi (2^{p_1-p_1'} (\hbox{sgn} (\xi_k)+\xi_k), 2^{p_2-p_2'}
(\hbox{sgn}(\zeta_k)+\zeta_k)). 
\eeq

By (\ref{support}) and the definition of $\hat\psi$ it is clear that
$\Phi_{k,l}$ are zero if $\bp\neq \bp'$. Consequently  the sensing matrix
is the block-diagonal matrix with each block (indexed by $\bp=\bp'$)
in the form of random Fourier matrix 
\beq
\Phi_{k,l}= {1\over {2n_\bp+1}} e^{i\pi (q_1\xi_k +q_2\zeta_k)}. 
\eeq

Let  $X=(X_l)$ with
\[
X_l=2\pi {(2n_{\bp}+1) 2^{(p_1+p_2)/2}}\nu_{\bp,\bq}
\]
be the target vector. Let 
\[
m=\sum_{j_1=-p_*}^{p_*}\sum_{j_2=-p_*}^{p_*}
(2m_\bj+1)^2. 
\]
We can then write the measurement vector $Y=\bA X$ where $\bA\in \IC^{n\times m}$.
The above observation  means that  the target structures of different dyadic scales are decoupled and can be determined  separately by our
approach using compressed sensing techniques.

\commentout{
Let us fix the powers $\bp=(p_1, p_2)\in \IZ^2$ of the dyadic scales. Without loss of generality, we may assume $p_1\geq p_2$.   Denote the target on the scales $(2^{p_1}, 2^{p_2})$
by 
\[
X_\bp=2\pi{(2n_{\bp}+1) 2^{|\bp|_1/2}} \lt(\nu_{\bp,\bq}\rt)_{|\bq|\leq m_\bp},
\]
the best $s$-term approximation  of $X_\bp$ by $X_\bp^{(s)}$ 
and the BP-estimator by  $\hat X_\bp$. 
 Let us generalize  Schemes I and II to the case 
with (\ref{sense2}) which is the anisotropic version of random
Fourier matrix  and  hence does not possess the square-symmetry.
}

 To solve (\ref{200'})-(\ref{201'}) we consider 
\beq
\label{ch1''}
\theta_k+\tilde\theta_k=2\phi_k+\pi
\eeq
 and 
\beq
\label{ch2'}
2\sin{\theta_k-\tilde\theta_k\over 2}=\rho_k.
\eeq
To  satisfy the relationships  (\ref{ch1''}) and (\ref{ch2'}) we set  
\beq
\label{46}
\theta_k&=&\phi_k+{\pi\over 2}+\arcsin{\rho_k\over 2}\\
\tilde\theta_k&=&\phi_k+{\pi\over 2}-\arcsin{\rho_k\over 2}
\label{47}
\eeq
analogous to (\ref{21}) and (\ref{22}). 

In the case of extreme anisotropy (needle-like structure),  say  $2^{p_1-p_2}\gg 1$,  (\ref{40})-(\ref{40'}) implies
\[
\phi_k\approx \pm {\pi\over 2}
\]
and hence 
\beq
\label{300}
\theta_k+\tilde\theta_k\approx 0
\eeq
by (\ref{ch1''}). 
On the other hand if $2^{p_1-p_2}\ll 1$ then
\[
\phi_k\approx \pm \pi
\]
and hence 
\beq
\label{301}
\theta_k+\tilde\theta_k\approx \pi.
\eeq
Relations (\ref{300}) and (\ref{301})  are reminiscent of Snell's law of reflection
if $\tilde \theta_k$ are interpreted as the reflection angles.

\commentout{
\bigskip

\noindent{\bf Scheme I'.}
Consider again  the backward sampling  condition  (\ref{41}).

When   $2^{p_1-p_2}\gg 1$
(\ref{41}) and (\ref{300}) imply that
$\theta_l \approx -\tilde\theta_l \approx \pi/2$. That is,
Scheme I dictates  the incident and sampling directions to be
nearly perpendicular to the direction 
of the larger scale in the case of extreme anisotropy. 

\bigskip

\noindent {\bf Scheme II'.}
Let 
\beq
\label{43'}
\om_l={\pi \gamma\over \sqrt{2^{2p_1} +2^{2p_2}}}
\eeq
with the constant
\[
\gamma={\rho_l\over \sqrt{2} \sin{\theta_l-\tilde\theta_l\over 2}}\geq 1
\]
and choose the incident and sampling angles according to 
\beq
\label{21'}
\theta_l=\phi_l+{\pi\over 2} +\arcsin{\rho_l\over \gamma\sqrt{2}}\\
\tilde\theta_l=\phi_l+{\pi\over 2}-\arcsin{\rho_l\over \gamma\sqrt{2}}.\label{22'}
\eeq
As before, in the high frequency limit $\gamma\to \infty$, 
the forward sampling condition holds, i.e.
$\tilde\theta_l\approx \theta_l$ which in the case of extreme 
anisotropy (\ref{300}) implies that
$\tilde\theta_l\approx \theta_l\approx 0.$
Namely, in the high frequency limit Scheme II dictates 
the incident and sampling directions to be
nearly parallel to the direction of the larger scale
of an extremely  anisotropic target structure.   
}

Using the RIP for the random Fourier matrix of each block,
we obtain the following theorems analogous to Theorem \ref{thm:mf}. 

\begin{theorem} \label{thm3}
 For each $\bp, |\bp|\leq p_*$, let (\ref{77}) be satisfied for  $n=n_\bp$ and $\sigma=2s_\bp$ and 
 any $\delta<\sqrt{2}-1$. Suppose
 \beq
 \label{freq2}
 \om_k\geq 2\pi \cdot 2^{\max{(p_1, p_2)}},\quad k=1,...,n_\bp.
 \eeq

Let $\xi_l, \zeta_l$ be independently and
uniformly distributed in $[-1,1]$. Let
the incident and sampling  angles 
be determined by (\ref{40})-(\ref{40'}), (\ref{200'})-(\ref{201'}). 
 
  Then (\ref{ric}) is satisfied 
for the $\bp$-block of the sensing matrix 
with probability at least $1-\alpha$. Consequently
the solution $\hat X_{\bp}$ by BP (\ref{32})  satisfies
 \beq
% \label{100'}
% \|\hat X_{\bp}-X_{\bp}\|_1&\leq & C_0\|X_{\bp}-X^{(s_\bp)}_{\bp}\|_1\\
 \|\hat X_{\bp}-X_{\bp}\|_2&\leq & C_1s_\bp^{-1/2}\|X_\bp-X^{(s_\bp)}_\bp\|_1+C_2{\ep}\label{101'}
 \eeq
 for the same constants $C_1, C_2$ as in (\ref{101}).

\end{theorem}
\begin{remark}
 The  condition (\ref{freq2}) implies
that the wavelengths are no larger than the scales
under interrogation, consistent  with the classical resolution criterion.

Theorem \ref{thm3} allows reconstruction  with probes of single 
sufficiently high frequency $\om_k=\om^\#$
\[
\om^\#\geq \pi 2^{1+p_*}.
\]
The beauty of the theorem, however, lies in the fact
that with probes of any frequency $\om$ the
coefficients in the Littlewood-Paley expansion for
scales up to $\om/(2\pi)$ can be  recovered (exactly in the
absence of noise). 

\end{remark}

\commentout{
\begin{theorem}\label{thm8'}
Let $Y_\bp$ be contaminated by noise of size $\ep$ and  let $\hat X_\bp$ be the corresponding solution to (\ref{32}). Under  the assumptions of Theorem \ref{thm3},
we have
 \beq
 \label{11}
 \|\hat X_\bp-X_\bp\|_2\leq C_1s_\bp^{-1/2} \|X_\bp-X^{(s_\bp)}_\bp\|_1+ C_2\ep
 \eeq
 with
probability at least $1-\alpha$
where  $C_1$ and $C_2$ are
constants. 
    \end{theorem}
}

\commentout{% With isotropic filter function
\section{Localized  extended targets}\label{sec:ext}
In this section we consider  localized  extended targets
and represent them by interpolating from grid points.

Suppose that the target function $\nu(\br)$ is continuous 
and has  a compact support. 
\commentout{ We write 
\beq
\label{530}
\nu(\br)=\int \delta_D(\br-\br') \nu(\br') d\br'
\eeq
where $\delta_D(\br)$ is the Dirac-delta function
and  consider the filtered discretization of (\ref{530}) as follows.
%Clearly $g_\eta$ tends to $\delta$ as $\eta\to 0$. 
}
Consider the filtered version $\nu_\eta$ of  $\nu$
\beq
\label{531}
\nu_\eta(\br)=\int g_\eta(\br-\br') \nu(\br') d\br'
\eeq
where  $g_\eta(\br)=
\eta^{-2} g(|\br|/\eta)$ for any integrable filter  function $g$ such that $\int g(\br) d\br =1$. 
Clearly, $\nu_\eta$ tends to $\nu$ as  $\eta\to 0$.
%Clearly, $\|v_\eta-v\|_\infty\to 0$ as $\eta\to 0$. Moreover,
%for square-integrable $v$, $v_\eta$ tends to $v$ in $L^2$
%as $\eta \to 0$.  The size of $\eta$ will be determined later.

Next we discretize (\ref{531}) by replacing the integral by
the Riemann sum of step size $\ell$. We obtain
\beq
\label{533}
\nu_{\eta,\ell}(\br)=\ell^2\sum_{\bp\in I}
g_\eta(\br-\ell\bq ) \nu(\ell\bq),\quad I\subset \IZ^2 
\eeq
which is as smooth as the interpolation element $g_\eta$ is. 
%For sufficiently small $\eta, \ell$,
%$\|v_{\eta, \ell}-v\|_{1}$ can be made arbitrarily small. 
\commentout{ the difference in
the Sobolev norm $\|v_{\eta, \ell}-v\|_{k,1}$ can be
made arbitrarily small where
\[
 \|v\|_{k,1}=\max_{j\leq k} \lt\|{\partial^k\over \partial x^j\partial z^{k-j}} v \rt\|_1,\quad k\in \IN. 
 \]
 }

Since $\nu$ has a  compact support, $I$ is a finite set. 
For simplicity let  $I$ be the square sublattice 
\[
I=\{\bq=(q_1, q_2): q_1, q_2= 1,...,\sqrt{m}\}
\]
of total cardinality $m$. Let $j=(q_1-1)\sqrt{m}+ q_2$. 
Define the target vector $X=(X_j)\in \IC^m$ with $X_j=\nu(\ell\bq)$. Let $\om_l$ and $ \bd_l$ be the probe frequencies
and directions, respectively, and let $\hat\br_l$ be the
sampling directions for $l=1,...,n$. 
Now we write
the data vector $Y$ 
in the form (\ref{31}) with  the sensing matrix elements 
\beq
\label{535}
\Phi_{lj}&=&{1\over  2\pi  \hat g_\eta(\om_l|\bd_l-\hat\br_l|)}
\int_{\IR^2} g_\eta(\br'-\ell\bq) e^{i\om_l (\bd_l-\hat\br_l)\cdot \br'} d\br',\quad j=(q_1-1)\sqrt{m}+ q_2\\
&=& e^{i\om_l \ell \bq\cdot (\bd_l-\hat\br_l)}\nn
\eeq
and the error term $E$ due to the filtered discretization. 
For sufficiently small $\eta, \ell$ we may assume $\|E\|_2\leq \ep n^{1/2}$ for a given $\ep>0$ (see Remark \ref{rmk1} and below). 

The crucial  observation is that the sensing matrix (\ref{535}) is
identical to (\ref{entry}) with $(x_j, z_j)=\om\bq$ and
any {\em isotropic} filter function $g$.
Therefore
Theorem \ref{thm:mf} holds  verbatim  for
 the case of localized  extended targets formulated
above. 

How small must $\eta$ and $\ell$ be in order to ensure
that $\|E\|_2\leq \ep n^{1/2}$? This  
can be answered roughly as follows. First, by the inequality $\|E\|_2\leq \|E\|_\infty \sqrt{n}$
it suffices to have $\|E\|_\infty \leq \ep $. Now
consider the transformation
$\cT$, defined by 
\[
 (\cT \nu)_l ={1\over 2\pi \hat g_\eta(\om_l|\bd_l-\hat\br_l|)} \int \nu(\br') e^{i\om_l (\bd_l-\hat \br_l ) \cdot \br'} d\br', 
 \]
 from  the space of continuous functions supported on  $[\ell, m\ell]^2$ to
$\IC^n$,  cf. (\ref{sa}). By definition
  \[
 E= \cT v -\cT v_{\eta,\ell}
 \]
 and we have
 \[
 \|E\|_\infty\leq {\|\nu-\nu_{\eta,\ell}\|_1\over  2\pi \min_{l}|\hat g_\eta(\om_l|\bd_l-\hat\br_l|)|}.
 \]
 
 %\[
%g_\eta(\br)={1\over {2\pi}\eta^2} e^{-{|\br|^2\over 2\eta^2}}. 
%\]
For  the sampling schemes I and II (with $\gamma=1$)  of Section \ref{sec:mf}, we can give an explicit bound for
the discretization error
\beqn
\|E\|_\infty\leq {1\over 2\pi}  {\|\nu-\nu_{\eta,\ell}\|_1}
\|\hat g^{-1}\|_{L^\infty([0, 2\eta \Om])}. 
\eeqn
where $\|\cdot\|_{L^\infty([0, 2\eta\Om)}$ denotes
the $L^\infty$-norm of functions defined on $[0, 2\eta \Om]$. 
 
 \begin{theorem} \label{thm:loc-ext}
Consider the sampling schemes I and II (with $\gamma=1$)  of Section \ref{sec:mf}. Assume  (\ref{75})  and
\beq
\label{disc-err}
{1\over 2\pi}  {\|\nu-\nu_{\eta,\ell}\|_1}
\|\hat g^{-1}\|_{L^\infty([0, 2\eta\Om])}\leq \ep. 
\eeq
Suppose that (\ref{77}) 
 holds with $\sigma=2s$ and any $\delta<\sqrt{2}-1$.   Then 
the bound (\ref{101})
holds  true  with $\kappa=n$ and with probability at least $1-\alpha$.
%In particular, every target vector of sparsity less than $s$
%can be exactly recovered  by  BP (\ref{L1}).
\end{theorem}

\begin{remark}
To give an explicit example, consider
 \[
g(|\br|)={1\over {2\pi}} e^{-{|\br|^2\over 2}} 
\]
with
$
\hat g(|\bk|)=g(|\bk|).$ 
With the choice $\eta=\ell$, the condition (\ref{disc-err}) can be formulated as
\[
  {\|\nu-\nu_{\ell,\ell}\|_1}\leq
 e^{\pi^2} \ep. 
 \]

\end{remark}
}
\commentout{% another paper: Compressively computed tomography
\section{Application: Computed tomography (CT)}
Theoretically, all reconstruction algorithms in  computed tomography (CT)
may be regarded as methods for inverting 
the Radon transform. However, they are by no means
equivalent in practice due to various approximations in implementation. For example, 
for the method of direct Fourier
reconstruction the dominant error is the interpolation error between the polar (sampling) grid
and the rectangular (reconstruction)  grid in the Fourier domain  \cite{Nat}.
 The purpose of this section is to point out  that the sampling scheme (\ref{freq}),
(\ref{46})-(\ref{47}) may be used, after slight modification,   for  {\em compressive  inversion}  of   the Radon
transform which has the additional benefit of
avoiding the interpolation error of 
the direct Fourier reconstruction. 
%For definiteness, we set
%\[
%\om_k=2\pi \cdot 2^{-\min{(p'_1, p'_2)}}.
%\]

Let us begin by assuming that the unknown function to be reconstructed
from the Radon transform is given by (\ref{30}).
%with only a finite number of nonzero coefficients. 
Let $\mu$ be the Radon transform of $\nu$:
\[
\mu(\hat \br, s)=\bR\nu(\hat\br,s)\equiv \int_{\hat\br\cdot\br'=s} \nu(\br') d\br'.
\]
It is easy to see that 
\[
\mu(\hat \br, s)={1\over 2\pi} \int\int \nu(\br') e^{i\om (s-\hat \br \cdot\br')} d\br' d\om. 
\]

 Let $\alpha_k,\beta_k$ be given as in (\ref{40}) and (\ref{40'}) and let $(\rho_k,\phi_k)$ be the polar coordinates
 of $(\alpha_k,\beta_k)$  where $k$ is
related to $(\bp',\bq')$ as before.   Define $\hat\br_k=(\cos\phi_k,\sin\phi_k)$. 
Let $G_{n,N}$ be the random polar grid 
\[
G_{n,N}=\{(\hat \br_k, s_l):  k=1,...,n, l=-N,...,N-1\}
\]
where $s_l=hl$ with $h>0$ being the step size in the radial direction. Let  $\mu$ be sampled at $G_{n,N}$. 

The  reconstruction scheme consists of three
steps:

{\em Step 1:} For $k=1,...,n$ compute the discrete approximation 
\beq
\label{152}
\tilde \mu(\hat\br_k, \om_k) ={h\over \sqrt{2\pi} } \sum_{l=-N}^{N-1} 
e^{-i\om_k s_l }\mu(\hat\br_k, s_l)
\eeq
to the  Fourier
transform of $\mu$ in the radial direction
\beq
\label{153}
\cF_\rho \mu(\hat\br_k, \om_k) 
= {1\over \sqrt{2\pi}} \int \nu(\br') e^{-i\om_k\hat \br_k \cdot\br'} d\br'. 
\eeq

This step is the same as in the  Fourier reconstruction
\cite{Nat}. The resulting error is due to  discretization. 

{\em Step 2:} Reconstruct $\nu$ from (\ref{153}) with
the approximate data (\ref{152}) by the compressed sensing
technique. By (\ref{30}), the sensing matrix in this case is 
given by
\beq
\label{155}
\Phi_{k,l}= {1\over {2n_\bp+1}}\hat\psi(\om_k 2^\bp
\hat\br_k)
 e^{-i\om_k 2^{\bp}\hat\br_k\cdot \bq}
\eeq
where $l$ is related to $(\bp, \bq)$ as before. 
By the same analysis, the sensing matrix 
is block-diagonal with each block in  the form
of random Fourier matrix satisfying RIP. Apply the
convex relaxation (\ref{32}) to obtain 
the approximate $(\tilde \nu_{\bp, \bq})_{\bp,\bq\in \IZ^2}$ and
form 
\[
\tilde\nu (\br) =\sum_{\bp,\bq} \tilde\nu_{\bp,\bq} \psi_{\bp,\bq}(\br).
\]

This step replaces the step of  interpolating the rectangular
grid from 
the polar grid and the final step of discrete inverse Fourier transform 
in the Fourier reconstruction \cite{Nat}. 

}

 \section{Numerical results}
 \label{sec:num}
     \begin{figure}[t]
\begin{center}
\includegraphics[width=0.8\textwidth,height=7cm]{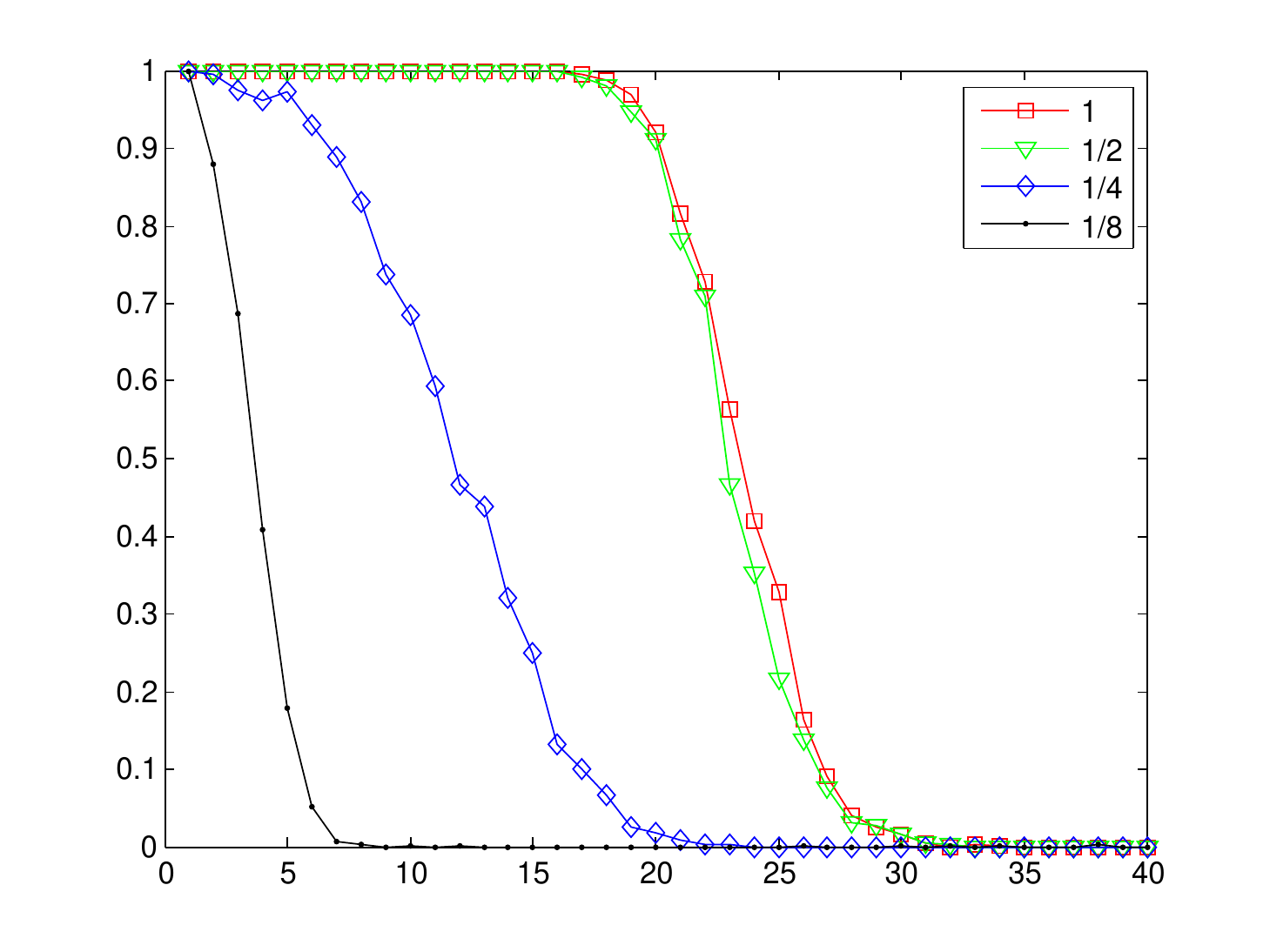}
\end{center}
\caption{Success probabilities for Scheme I with (\ref{30'}). The label indicates the value of $\eta$. As the recipe (\ref{41}) is increasingly violated, the performance degrades accordingly. }
\label{fig1}
\end{figure}

    \begin{figure}[t]
\begin{center}
\includegraphics[width=0.8\textwidth, height=7cm]{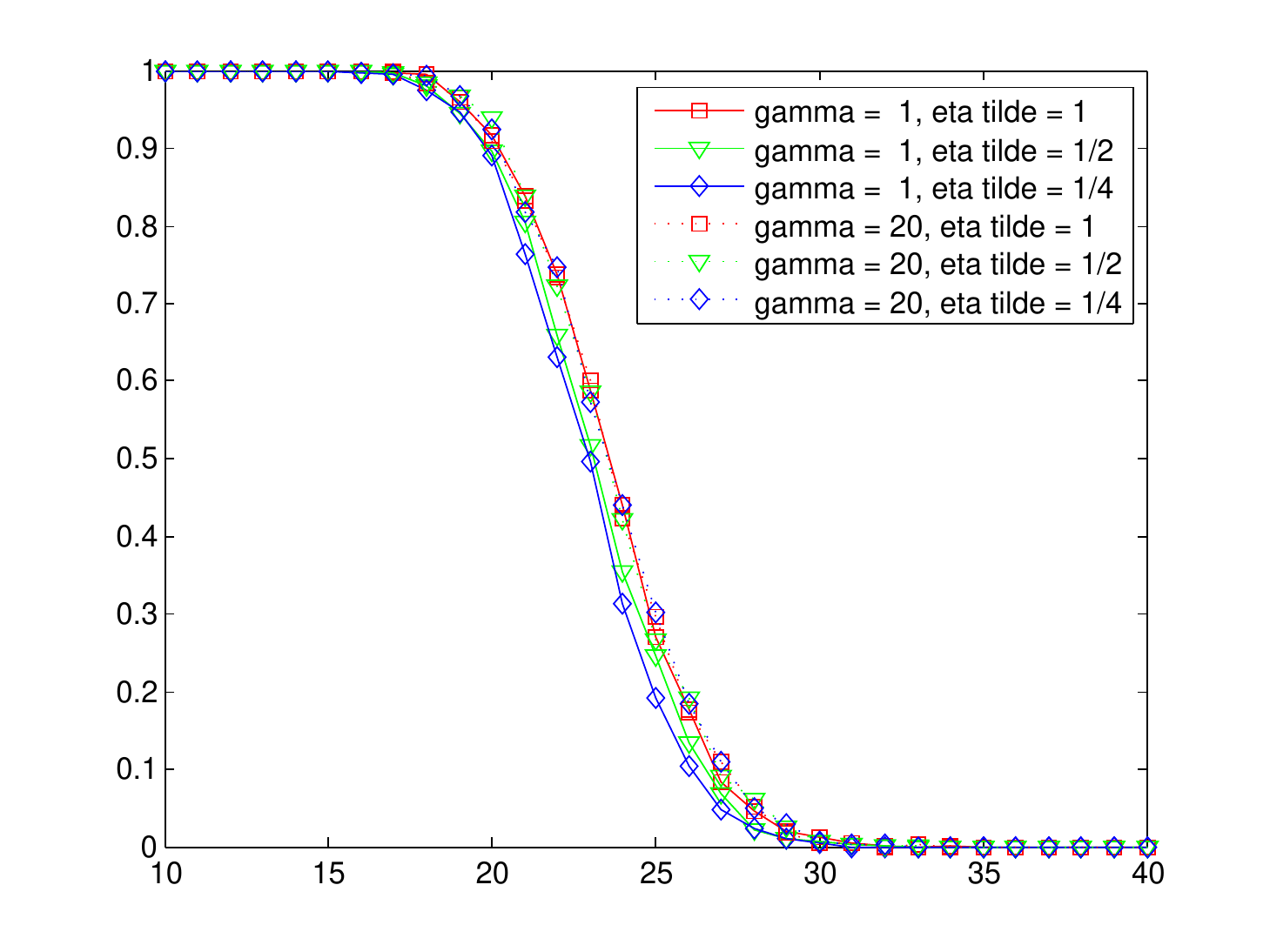}
\end{center}
\caption{Success probabilities for Scheme  II (a): $\eta=1, \tilde\eta=1,1/2, 1/4$ with $\gamma=1, 20$.}
\label{fig3}
\end{figure}

    \begin{figure}[t]
\begin{center}
\includegraphics[width=0.8\textwidth, height=7cm]{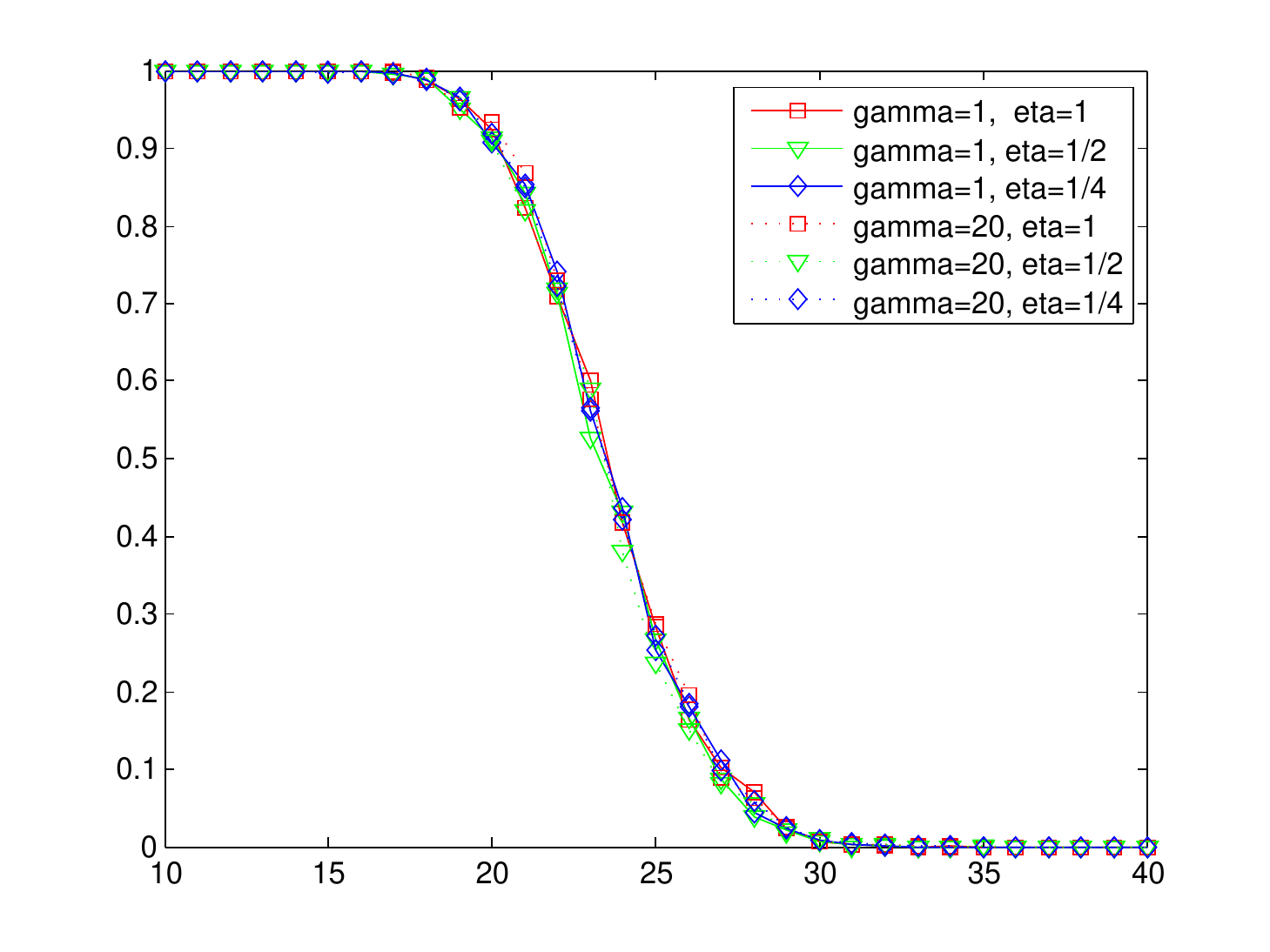}
\end{center}
\caption{Success probabilities for Schemes  II (b): $\eta=\tilde\eta=1,1/2,1/4$ with $\gamma=1,20$.} 
\label{fig4}
\end{figure}

    \begin{figure}[t]
\begin{center}
\includegraphics[width=0.8\textwidth,height=7cm]{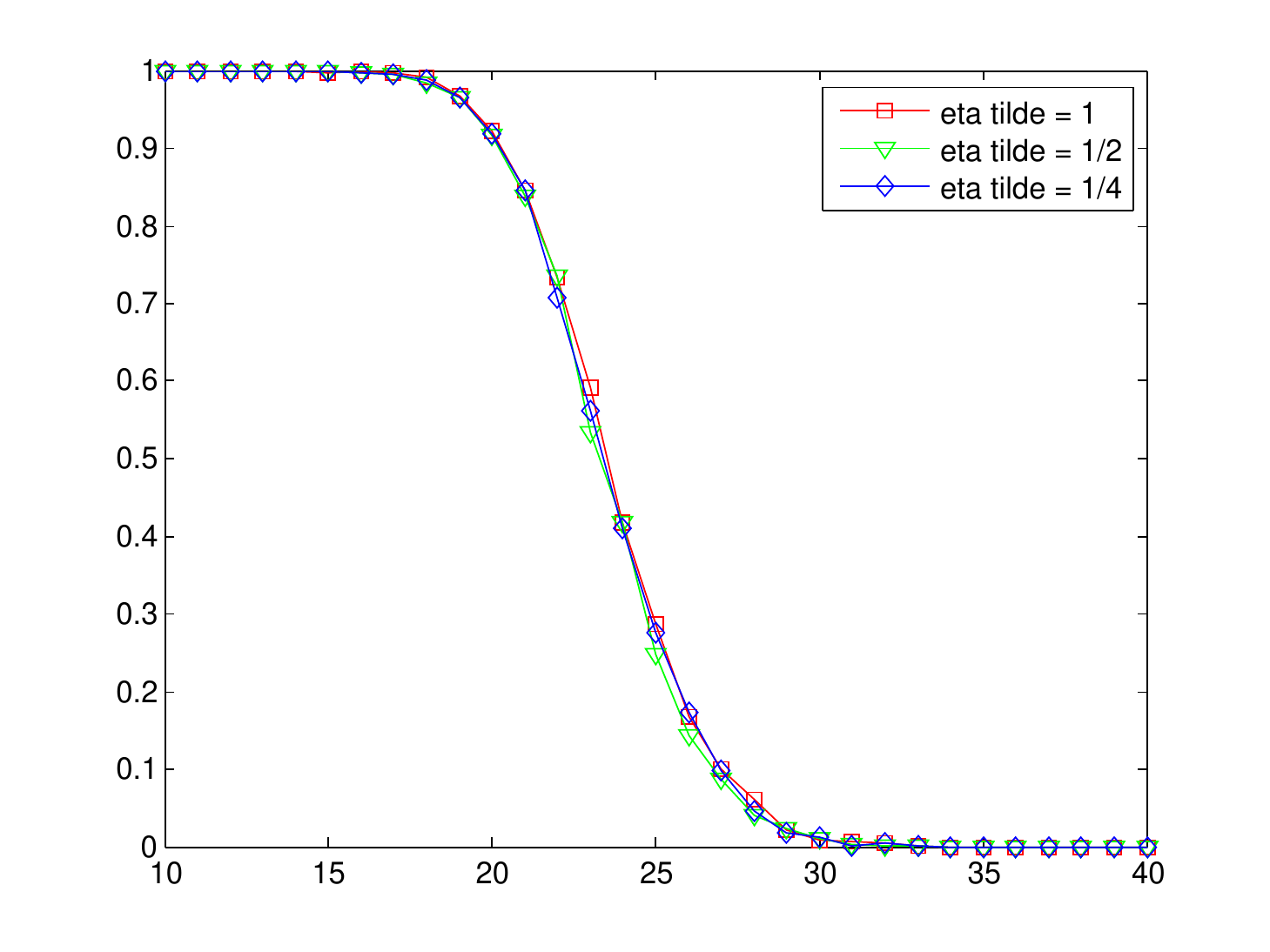}
\end{center}
\caption{Success probabilities for Scheme  III
with (\ref{23-1}). } 
\label{fig5}
\end{figure}

\begin{figure}[t]
\begin{center}
\includegraphics[width=0.8\textwidth, height=7cm]{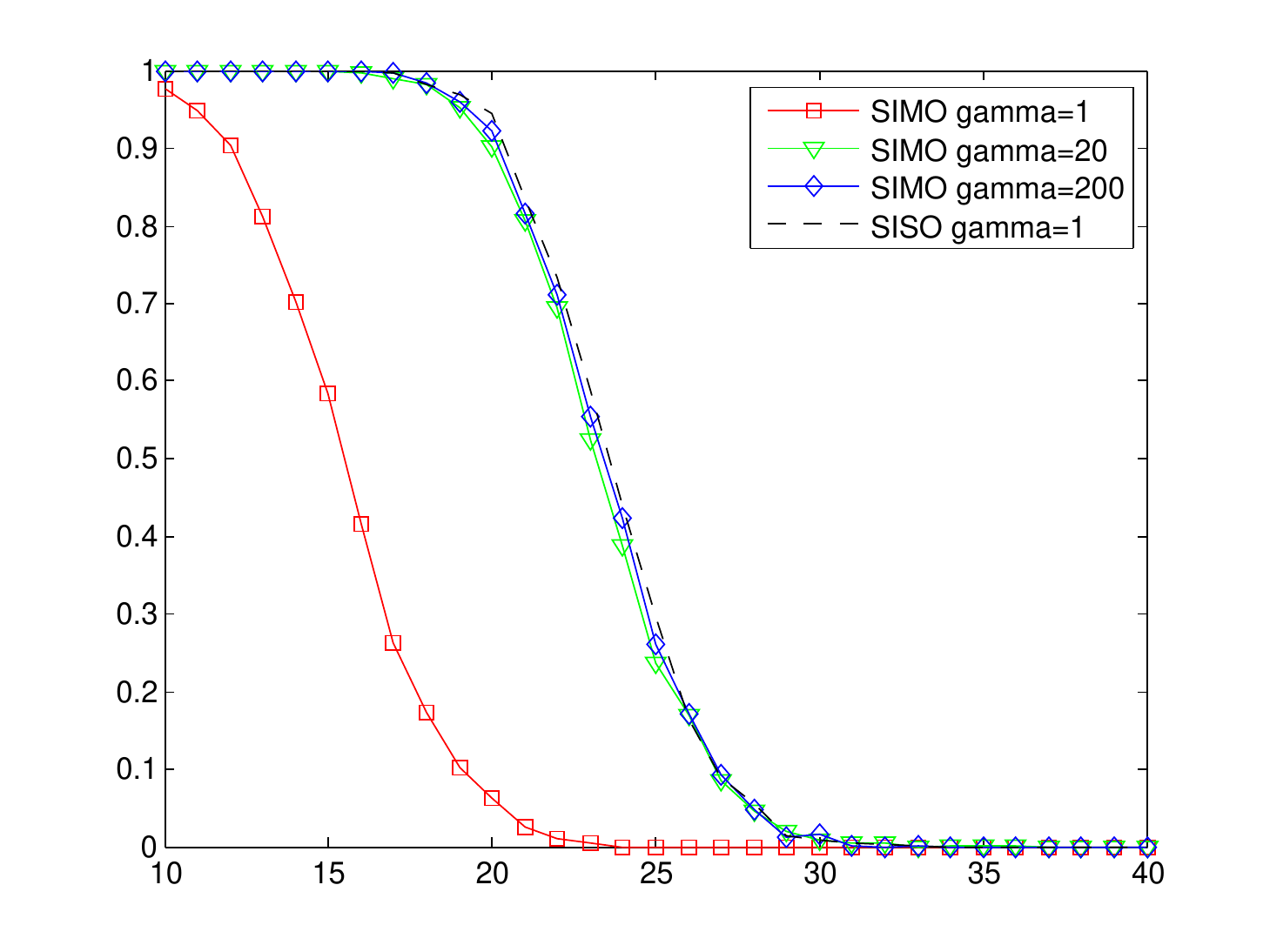}
\end{center}
\caption{Solid curves are the success probabilities
for the SIMO measurement at $\gamma=1,20,200$ and
the dashed curve is the SISO Scheme II at $\gamma=1$. }
\label{fig6}
\end{figure}
Greedy algorithms have significantly lower  computational
complexity than linear programming and have 
provable performance under various conditions.
For example
under the condition  $\delta_{3s}<0.06$
 the Subspace Pursuit (SP) algorithm is guaranteed to exactly recover $X$ via a finite number of iterations \cite{DM}.
 We have used SP for  reconstruction  in all our  simulations
 with  
 the following parameters: $m = 2500,   \ell = 1, \Omega = \pi/\sqrt{2}, n = 100$. The probability of recovery is calculated
 by using 1000 independent runs. 
 In Figures \ref{fig1}-\ref{fig6}, the vertical axis is
 for the probability of recovery and the horizontal axis
 is for the number of point scatterers.

To test  Scheme  I numerically, we use (\ref{42})
and 
\beq
\label{30'}
\tilde\theta_l=\theta_l+\eta\pi,\quad \eta=1, 1/2, 1/4, 1/8
\eeq
 to see if the
deviations from (\ref{41}) have any impact on performance.
Their  probabilities of recovery are  plotted as
a function of the sparsity in Figure \ref{fig1}.
Clearly, the performance deteriorates rapidly as the
difference between the sampling and incident
angles decreases. This is due to  increasingly more frequent and
more severe violation of (\ref{const}) as a result.
 In other words,  the backward scattering
direction is the optimal sampling direction for Scheme I.

 Likewise, to test Scheme II numerically, we use
 (\ref{20}), 
 \beq
\tilde\theta_l&=&\phi_l+{\tilde\eta\pi\over 2}-\arcsin{\rho_l\over \gamma\sqrt{2}},\quad \tilde\eta=1, 1/2, 1/4, 
\label{21-1} 
\eeq 
instead of (\ref{22}), 
and (\ref{21})  with (a) $\eta=1$ as well
as  with (b) $\eta=\tilde\eta=1, 1/2, 1/4.$
Condition (\ref{25}) is satisfied if and only if
$\eta=\tilde\eta$. 

The results for
case (a)  at $\gamma=1,20$
are shown in
Figure \ref{fig3} and the results for case (b) with $\eta=\tilde\eta$  at $\gamma=1, 20$ are displayed in  Figure \ref{fig4}. 
The slight degradation in performance for e.g., $\eta=1, \tilde\eta=1/4, \gamma=1$ results from the violation of (\ref{25})
(and hence (\ref{const})) which affects the performance 
significantly  for small $\gamma$. 
For  large $\gamma$, it has been shown in \cite{cis-simo} that 
the sampling angles can be chosen independently of
the incident angles to maintain a high level of performance 
(Figure \ref{fig6}). 
On the other hand, when  (\ref{25})  holds, 
the performance is essentially independent of both 
$\gamma$ and  $\tilde\eta$,  Figure \ref{fig4}.

For Scheme III  we use
 equally spaced incident angles $\theta_l\in [\pi/6, \pi/3]$, 
 (\ref{ch2}) and 
 \beq
 \label{23-1}
 \theta_l+\tilde\theta_l=2\phi_l+\tilde \eta \pi,\quad \tilde  \eta=1, 1/2, 1/4.  \eeq
 As shown in Figure \ref{fig5}, the performance is essentially
 independent of  $\tilde\eta$. 
 
We demonstrate numerically   the high-frequency
SIMO schemes  analyzed  in \cite{cis-simo}
and compare their performance  with that of the multi-shot SISO schemes presented above. 
A case in point    would
be to use  (\ref{20}) and (\ref{22}) but for  a fixed incident angle, say $\theta_l=0,\forall l$
(In this case (\ref{25}) is almost certainly  violated). 
In \cite{cis-simo} it is established that the  SIMO 
schemes with sampling angles, {\em independent of the incident angles}, achieves a high performance in reconstructing a sparse target with 
 a sufficiently high-frequency probe wave (i.e. $\gamma\gg 1$). 
The success probabilities of the SIMO schemes for $\gamma=1,20,200$
are calculated and plotted in Figure \ref{fig6}. 
Consistent with the theory \cite{cis-simo}, 
the low frequency case with $\gamma=1$ has
the worst performance. 
 Clearly the performance of the SIMO schemes
improves with $\gamma$ and in the limit $\gamma\gg 1$
approaches that of the multi-shot SISO Scheme II. 
There is negligible difference between the performances for $\gamma=20$ and $\gamma=200$ both of which follow closely
that 
 of the SISO Scheme II with $\gamma=1$ (black dashed line in Figure \ref{fig6}).
 
\begin{figure}[t]
\begin{center}
\includegraphics[width=0.8\textwidth, height=7cm]{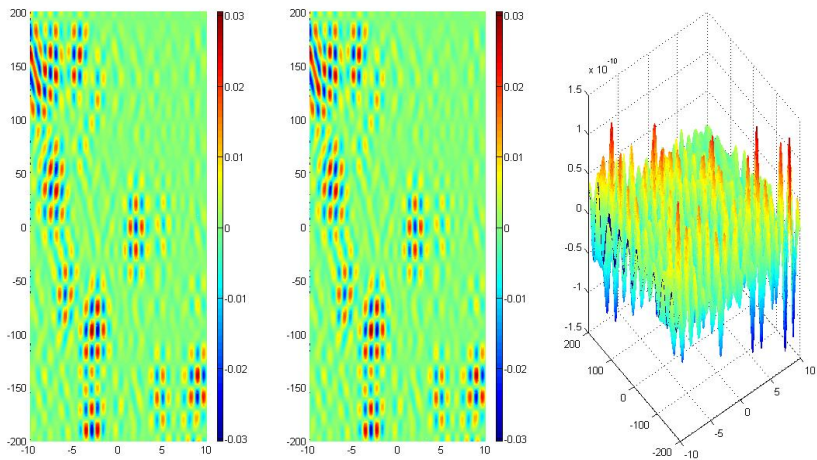}
\end{center}
\caption{Imaging of an extended target of the scales $(1,32)$: The left plot is the exact structure; the middle plot is the  the reconstructed profile; the right plot
shows the (round-off) error. See the color bars for the meaning
of  the color code.  }
\label{fig10}
\end{figure}

 Finally we demonstrate the reconstruction  with
 the Littlewood-Paley basis. Since the scattering matrix is block-diagonalized by the proposed sampling scheme (\ref{40})-(\ref{40'}) and (\ref{200'})-(\ref{201'}), we 
  consider  targets of single pair of dyadic scales. In this simulation, the reconstruction is carried out for 
a target on the scales $(1, 32)$ ($\bp=(0,5)$) with
  $42$ nonzero coefficients ($s_\bp=42$) among
 $441$ possible  modes ($ m_\bp=10$, $21$ modes in each coordinate) by using  $81$ samples ($n_\bp=81$). The target and
  its reconstruction in the domain $[-10,10]\times [-200, 200]$ are color-coded and displayed in Figure \ref{fig10}.

 \section{Conclusion and discussion}\label{sec:con}
 We have proposed, analyzed and numerically tested several  
 multi-shot SISO sampling schemes which transform
 the scattering matrix into the random Fourier matrix
 in the case of point and localized extended scatterers and 
the   block-diagonal  form of random Fourier matrices
 in the case
of distributed extended targets.  
 
 In the case of point  scatterers,  these sampling  schemes are
 either multi-frequency  band limited (I) or single frequency
 outside band (II).   
 For Scheme I , the sampling direction is the
 backward direction analogous to  synthetic aperture
 radar (SAR)  while for Scheme II in the
 high frequency limit the optimal
 sampling direction is in the forward direction, analogous
 to the X-ray tomography. 
Both schemes
 produce  nearly the  same recovery probability
  with the resolution given by (\ref{75}), i.e.
 \[
 \ell={\pi\over \sqrt{2} \Om}. 
 \] 
% when conditions (\ref{ch1}) and (\ref{ch2}) are satisfied. 

We have extended this approach to the case of
localized extended targets by interpolating
from grid points. We have formulated the approximate
scheme as inversion with noisy data. In particular,
we have derived an explicit  error bound for 
 the simple piece-wise constant
interpolation.  

In the case of distributed extended targets, the block-diagonal form
of the scattering matrix in the Littlewood-Paley representation 
means that different dyadic scales of the target are decoupled and can be imaged 
scale-by-scale separately by our method.
Moreover,  we can determine the
coefficients in the Littlewood-Paley expansion  (\ref{30}) for
scales up to $\om/(2\pi)$   by
using probes of any single frequency $\om$. 
 The disadvantage of the Littlewood-Paley
basis  is that  a localized target has  slowly
decaying coefficients and hence is not compressible in
this basis. 

\commentout{
 In the case of anisotropic extended targets,  as long as the target is comparable to wavelength in one direction,  
the target's subwavelength structures  in the other directions can
be resolved by our imaging methods.  This is an anisotropic superresolution
effect. 
}
 
 The SIMO schemes in which 
 the scattered field of an incident  angle is measured at multiple
 sampling angles 
 have  been studied in \cite{cis-simo,  subwave-cs, cs-par}. 
 Except for the special case of the periodic scatterers lying on a transverse plane \cite{subwave-cs}, it is not known  if the sensing matrices of the SIMO schemes in general satisfy the RIP or not. In this case, the  approach based on the notion of incoherence is taken  to 
analyze the  SIMO schemes \cite{cis-simo, cs-par}. 
This  approach is generally more flexible and should be applicable 
to the SISO schemes considered here. 
%Because the RIP approach,
%when it works, works for all { targets} satisfying the sparsity constraint. 

The main advantage of the  SIMO schemes is that
in the one-shot setting (one incident field) the inverse
scattering problem can be solved exactly {\em without} the Born approximation by inverting an auxiliary nonlinear system of
equations \cite{cis-simo}. We are working to extend the idea to the  imaging methods with multi-shot measurements. 

On the other hand, the SISO schemes with the RIP tend to have a better
performance which can be matched by that of the SIMO schemes
only at high frequency as demonstrated  in Figure \ref{fig6}.
In the high frequency regime the sampling angles can
be chosen independently  of the incident angles \cite{cis-simo}.  

 \bigskip
 {\bf Acknowledgement.} I thank my students Hsiao-Chieh Tseng
 (Figures \ref{fig1}-\ref{fig6}) and Wenjing Liao (Figure \ref{fig10}) 
 for preparing the figures.

\end{document}